\documentstyle[aps,amssymb,preprint,epsfig,eqsecnum,floats,cite]{revtex}
\tighten
\def \be{\begin{equation}}
\def \ee{\end{equation}}
\def \bea{\begin{eqnarray}}
\def \eea{\end{eqnarray}}
\def \ben{\begin{enumerate}}
\def \een{\end{enumerate}}
\def \bit{\begin{itemize}}
\def \eit{\end{itemize}}
\def \branch{{\cal B}}

\def \Im{{\text{Im}}\,}
\def \Re{{\text{Re}}\,}
\def \sd{\text{SD}}
\def \GeV{{\text{GeV}}}

\def \MeV{{\text{MeV}}}
\def \TeV{{\text{TeV}}}
\def \av#1{\left\langle #1\right\rangle}
\def \B{\bar{B}}
\def \bm{\boldmath}
\def \braket#1#2#3{\langle #1|#2| #3\rangle}
\def \cl#1{{#1\%\ \mathrm{C.L.}}}

\def \cp{\mathrm{CP}}
\def \cpt{\mathrm{CPT}}
\def \diag{{\mathrm{diag}}}
\def \dis{\displaystyle}
\def \ea{{\it et al.}}
\def \eq#1{Eq.~(\ref{#1})}
\def \eqs#1#2{Eqs.~(\ref{#1})--(\ref{#2})}
\def \fig#1{Fig.~\ref{#1}}

\def \hadron{{\mathrm {had}}}
\def \hc{\mathrm{H.c.}}

\def \mgut{M_{\mathrm{GUT}}}
\def \nnu{\nonumber}
\def \ol#1{\overline{#1}}

\def \rf{Ref.~\cite}
\def \rfs{Refs.~\cite}
\def \Sec#1{Sec.~\ref{#1}}
\def \sm{\mathrm{SM}}
\def \tanw{\tan\theta_W}
\def \vckm{V_{\mathrm{CKM}}}

\def \chargino{\tilde{\chi}^{\pm}}
\def \charginoi#1{\tilde{\chi}^{\pm}_{#1}}

\def \gluino{\tilde{g}}
\def \higgs{H^{\pm}}
\def \neutralino{\tilde{\chi}^0}
\def \neutralinoi#1{\tilde{\chi}^0_{#1}}
\def \sleptoni#1{\tilde{l}_{#1}}

\def \squark{\tilde{q}}
\def \tsquark{\tilde{t}}
\def \tsquarki#1{\tilde{t}_{#1}}
\def \squarki#1{\tilde{q}_{#1}}
\def \squarkui#1{\tilde{u}_{#1}}
\def \squarkdi#1{\tilde{d}_{#1}}
\def \sneutrino{\tilde{\n}}
\def \sneutrinoi#1{\tilde{\n}_{#1}}
\def \R#1#2{r^{#1}_{#2}}
\def \rcontcc{R_{\mathrm {cont}}^{c\bar{c}}}

\def \rhadron{R_{\mathrm {had}}}

\def \rresrho{R_{\mathrm {res}}^{\r}}

\def \rrespsi{R_{\mathrm {res}}^{J/\psi}}
\def \cseff {c_7^{\text{eff}}}
\def \ceff {c_9^{\text{eff}}}
\def \ceffstar{c_9^{\text{eff}*}}
\def \a{\alpha}
\def \b{\beta}

\def \g{\gamma}
\def \G{\Gamma}
\def \d{\delta}
\def \epsi{\epsilon}

\def \k{\kappa}
\def \l{\lambda}

\def \m{\mu}
\def \n{\nu}

\def \p{\pi}
\def \r{\rho}
\def \s{\sigma}

\def \t{\tau}
\def\acta#1#2#3{{Acta Phys.~Pol. B} {\bf #1}, #3 (19#2)}
\def\ann#1#2#3{Annu. Rev. Nucl. Part. Sci. {\bf#1}, #3 (19#2)}
\def\euro#1#2#3{{Eur. Phys. J. C} {\bf #1}, #3 (19#2)}

\def\ibid#1#2#3{\emph{ibid.} {\bf#1}, #3 (19#2)}
\def\ibidn#1#2#3{\emph{ibid.} {\bf#1}, #3 (20#2)}
\def\ib#1#2#3{{\bf#1}, #3 (19#2)}

\def\nc#1#2#3{{Nuovo Cimento A}~{\bf#1}, #3 (19#2)}
\def\np#1#2#3{{Nucl.~Phys.}~{\bf B#1}, #3 (19#2)}
\def\npn#1#2#3{{Nucl.~Phys.}~{\bf B#1}, #3 (20#2)}
\def\npps#1#2#3{{Nucl.~Phys.~B (Proc.~Suppl.)}~{\bf #1}, #3 (19#2)}
\def\pl#1#2#3{{Phys.~Lett. B}~{\bf #1}, #3 (19#2)}
\def\pr#1#2#3{{Phys.~Rev.}~{\bf #1}, #3 (19#2)}
\def\prd#1#2#3{{Phys.~Rev. D}~{\bf #1}, #3 (19#2)}
\def\prdn#1#2#3{{Phys.~Rev. D}~{\bf #1}, #3 (20#2)}
\def\prl#1#2#3{{Phys.~Rev.~Lett.}~{\bf #1}, #3 (19#2)}
\def\prln#1#2#3{{Phys.~Rev.~Lett.}~{\bf #1}, #3 (20#2)}
\def\prp#1#2#3{{Phys.~Rep.}~{\bf #1}, #3 (19#2)}
\def\ptp#1#2#3{{Prog. Theor.~Phys.}~{\bf #1}, #3 (19#2)}
\def\rmp#1#2#3{{Rev. Mod. Phys.} {\bf #1}, #3 (19#2)}
\def\zpc#1#2#3{{Z.~Phys. C}~{\bf #1}, #3 (19#2)}
\begin{document}
\preprint{\setlength{\baselineskip}{1.5em}
\vbox{\vspace{-1cm}
\hbox{FISIST/20-99/CFIF}
\hbox{hep-ph/0002089}
\hbox{February 2000}}}
\draft
\title{\Large\bf Flavour-Conserving \bm$\cp$ Phases in Supersymmetry and Implications for Exclusive \bm$B$ Decays}
\author{\sc 
F. Kr\"uger\thanks{E-mail address: krueger@gtae3.ist.utl.pt} 
and J.\,C. Rom\~ao\thanks{E-mail address: fromao@alfa.ist.utl.pt}}
\address{Centro de F\'{\i}sica das Interac\c{c}\~{o}es Fundamentais (CFIF),
Departamento de F\'{\i}sica,  \\ 
Instituto Superior T\'ecnico,  Av. Rovisco Pais,  1049-001 Lisboa, Portugal}
%
%
\maketitle
\begin{abstract}
We study rare exclusive $B$ decays based on the quark-level transition 
$b\to s (d)\, l^+l^-$, where $l=e$ or $\m$, in the context of 
supersymmetric 
theories with minimal flavour violation. 
We present analytic expressions for various 
mixing matrices in the presence of new $\cp$-violating phases, 
and examine their impact on observables involving 
$B$ and $\B$ decays.
An estimate is obtained for $\cp$-violating asymmetries in 
$\B\to K^{(*)}l^+l^-$ and $\B\to \r (\p)l^+l^-$ decays for the  
dilepton invariant mass region $1.2\, \GeV < M_{l^+l^-}< M_{J/\psi}$.
As a typical result, we find a $\cp$-violating partial width 
asymmetry of about $-6\%$ ($-5\%$) in the case of $B\to \p$ ($B\to \r$) 
in effective supersymmetry with phases of 
${\mathcal O}(1)$, taking 
into account the measurement of the inclusive 
$b\to s\g$ branching fraction. On the other hand, 
$\cp$ asymmetries of less than 
$1\%$ are predicted in the case of $B\to K^{(*)}$. We argue that it is not 
sufficient to have additional $\cp$ phases of ${\mathcal O}(1)$ to observe 
large $\cp$-violating effects in exclusive $b\to s (d)l^+l^-$ decays.    
\end{abstract}
%
%
\section{Introduction}\label{intro}
Within the standard model (SM), $\cp$ violation is caused by a 
non-zero complex phase in the Cabib\-bo-Ko\-ba\-ya\-shi-Maskawa 
(CKM) quark mixing matrix \cite{ckm}. While the experimentally 
observed indirect $\cp$ violation in the neutral kaon system, 
$\epsi_K$, can be accommodated in the SM, it is still an open question 
whether the SM description of $\cp$ violation is consistent with the 
new experimental result on direct $\cp$ violation, 
$\epsi'/\epsi_K$, 
since the theoretical prediction of its precise 
value suffers from large hadronic uncertainties \cite{epsilonprime}. 
On the other hand, if the baryon asymmetry of the universe 
has been generated via baryogenesis at the electroweak phase transition, 
the CKM mechanism of $\cp$ violation cannot account for the 
observed amount of baryon asymmetry.~This feature could be a hint of the 
existence of $\cp$-violating sources outside the CKM matrix \cite{baryogenisis}.

Important tests of the SM are provided by flavour-changing neutral current 
(FCNC) reactions involving $B$ decays \cite{ali:rev}, thus offering 
an opportunity to search for supersymmetric extensions of the SM 
\cite{joanne,susy:fcnc}. 
There are at present only a few FCNC processes which have been observed 
experimentally, but the situation will change 
considerably after the completion of $B$ factories in the near future.

In this work, we analyse the exclusive decays $\B\to K^{(*)}\,l^+l^-$ 
and $\B\to \r (\p)\,l^+l^-$ in the context of supersymmetry (SUSY) with 
minimal particle content and R-parity conservation 
\cite{mssm:original,mssmrev}.
The inclusive reaction $\B\to X_s l^+l^-$ 
within supersymmetric models has been extensively studied in 
\rfs{joanne,bertolini:etal,ali:etal,cho:etal,goto:etal,goto:etal:phase:sugra,off-diagonal,baek:ko,huang:etal} 
and, more recently, in \rf{lunghi:etal}. New physics effects in the exclusive 
channels have been investigated in \rfs{np1:exc,np2:exc,enrico:recent}.

We place particular emphasis on $\cp$-violating 
effects associated with the 
partial rate asymmetry between $B$ and $\B$ decays as well as the 
forward-backward asymmetry of the $l^-$. 
Within the SM these effects turn out to be unobservably small 
($\leqslant 10^{-3}$) in the decays $\B\to K^{(*)}l^+l^-$ \cite{four-body}, 
and amount to only a few per cent in $\B\to \r (\p)l^+l^-$ 
\cite{four-body,fklms:exc}. However, in models 
with new $\cp$-violating phases in addition to the 
single phase of the CKM matrix, larger effects may occur due to the interference of amplitudes with different phases. 
The purpose of the present analysis is to explore
$\cp$-violating observables in the aforementioned FCNC reactions that could  
provide evidence of a non-standard source of $\cp$ violation, and hence 
may be useful in analysing supersymmetry in future collider experiments.

The paper is organized as follows. In \Sec{mssm}, we exhibit
the various mixing matrices of the  minimal supersymmetric standard model
(MSSM) in the presence of additional $\cp$-violating phases. 
Within such a framework we discuss 
different scenarios for the SUSY parameters. 
In \Sec{rareBdecays}, we are primarily concerned with 
the short-distance matrix element and Wilson coefficients
governing $b\to s (d)\, l^+l^-$ in the MSSM. We also briefly describe 
an approximate procedure to incorporate quark antiquark resonant 
intermediate states -- namely $\r, \omega$, and the $J/\psi$ family --
which enter through the decay chain 
$b\to s(d)V_{q\bar{q}}\to s(d) l^+l^-$.
Section \ref{decays:exc} is devoted to the exclusive decay modes  
$\B\to K^{(*)}\,l^+l^-$ and $\B\to \p(\r)\,l^+l^-$, where 
formulae are given to calculate $\cp$ asymmetries
which can be determined experimentally by measuring the  
difference of $B$ and $\B$ events. 
In \Sec{results}, we present our numerical results for $\cp$-violating 
observables in the non-resonant domain $1.2\, \GeV < M_{l^+l^-}< M_{J/\psi}$,
taking into account experimental bounds on rare $B$ decays such as 
$b\to s\g$. We summarize and conclude in \Sec{conclusions}. 
The analytic formulae describing the short-distance effects in the 
presence of 
SUSY as well as the explicit expressions for the form factors are relegated 
to the Appendices. 

\section{The minimal supersymmetric standard model}\label{mssm}
In the MSSM, there are new sources of $\cp$ violation.
In general, a large number of $\cp$-violating phases appear in the 
mass matrices as well as the couplings. 
After an appropriate redefinition of fields one ends 
up with at least two new $\cp$-violating phases, besides the phase of the 
CKM matrix and the QCD vacuum angle, which cannot be rotated away.
For instance, in the MSSM with universal boundary conditions at some high 
scale only two new physical phases arise; 
namely $\varphi_{\m_0}$ associated with the Higgsino mass parameter $\m$ 
in the superpotential and $\varphi_{A_0}$ connected with 
the soft SUSY-breaking trilinear mass terms. 

In order to fulfil the severe constraints on the electric dipole moments 
(EDM's) of electron and neutron, one generally assumes that the new 
phases are less than ${\mathcal O}(10^{-2})$.
Since there is no underlying symmetry which would force the phases to be small,
this requires fine-tuning.
Of course, one can relax the tight constraint on these phases by having masses of the superpartners in the 
$\text{TeV}$ region; this heavy SUSY spectrum may, however, lead to an 
unacceptably large contribution to the cosmological relic density.    

It has recently been pointed out by several authors that it is 
possible to evade the EDM constraints so that phases of 
${\mathcal O}(1)$ still remain consistent with the current experimental 
upper limits. Methods that have been advocated to suppress the EDM's 
include cancellations among 
different SUSY contributions 
\cite{phase:canc,phase:non-uni}, and nearly degenerate heavy sfermions for the first two generations while being consistent with 
naturalness bounds. The latter can be realized within the context of so-called 
`effective SUSY' models \cite{effsusy}, thereby solving the SUSY FCNC and 
$\cp$ problems. 

To get an idea how supersymmetry 
affects $\cp$ observables in rare $B$ decays, we will consider as 
illustrative examples the following types of SUSY models:
\bit
\item MSSM coupled to $N=1$ supergravity with a universal 
SUSY-breaking sector at the grand unification scale.
%
%
\item Effective SUSY with near degeneracy of the heavy first and 
second generation sfer\-mi\-ons.
\eit
In the present analysis, we restrict ourselves to the discussion of
flavour-diagonal sfermion mass matrices -- that is, we assume the CKM matrix 
to be the only source of flavour mixing.\footnote{One should keep in mind 
that renormalization group effects induce flavour off-diagonal entries 
in the sfermion mass matrices at the weak scale (see below).}

\subsection{Mixing matrices and new \bm$\cp$-violating phases}
This subsection concerns the mass and mixing matrices relevant to our 
analysis. In what follows, we will adopt the conventions of 
\rf{mssmrev:romao}. 

\subsubsection{Charged Higgs-boson mass matrix}
The mass-squared matrix of the charged Higgs bosons reads
\be\label{massmatrix:higgs}
M_{\higgs}^2=
\left(\begin{array}{cc}B\m \tan\b + M_W^2 \sin^2\b + t_1/v_1 & B\m +M_W^2 \sin\b\cos\b\\ B\m + M_W^2 \sin\b\cos\b& B\m \cot\b + M_W^2 \cos^2\b + t_2/v_2
\end{array}\right),
\ee
with 
\be\label{bmu-term}
B\m=\frac{1}{2}\sin2\b (m_{H_1}^2 + m_{H_2}^2 + 2|\m|^2),
\ee
\be
|\m|^2= -\frac{1}{2}M_Z^2 + 
\frac{m_{H_2}^2\sin^2\b-m_{H_1}^2\cos^2\b}{\cos2\b}.
\ee
Here $B$ and $\m$ refer to the complex soft SUSY-breaking and Higgsino mass 
parameters respectively, $m_{H_{1,2}}^2$ are the soft 
SUSY breaking Higgs-boson masses at the electroweak scale, and   
$t_{1,2}$ stand for the renormalized tadpoles. The mixing angle $\b$ is 
defined as usual by $\tan\b\equiv v_2/v_1$, 
with $v_{1,2}$ denoting the tree level vacuum expectation values (VEV's) 
of the two neutral Higgs fields. 
In \eq{bmu-term} we have adjusted the phase of the 
$\m$ parameter in such a way that $B\m$ is real at tree level, 
thereby ensuring that the VEV's of the two Higgs fields are real.
Consequently, the mass matrix becomes real and can be reduced to a 
diagonal form through a biorthogonal transformation
$(M_{\higgs}^{\diag})^2=O M_{\higgs}^2 O^T$. At the tree level, i.e.~$t_i=0$ 
in \eq{massmatrix:higgs}, we have
\be
O=
\left(\begin{array}{cc}
-\cos\b& \sin\b\\\sin\b&\cos\b
\end{array}\right).
\ee
Before proceeding, we should mention that radiative corrections to the Higgs 
potential induce complex VEV's.
As a matter of fact, $\cp$ violation in the Higgs sector leads to an
additional phase which, in the presence of chargino and neutralino 
contributions, cannot be rotated away by reparametrization of fields 
\cite{veff:one-loop}. As a result, the radiatively induced phase 
modifies the squark, chargino, and neutralino mass matrices. 
In the present analysis, we set this phase equal to zero.

\subsubsection{Squark mass matrices}
We now turn to the $6\times 6$ squark mass-squared matrix which can be 
written as
\be\label{squark:massmatrix}
M^2_{\squark}=
\left(\begin{array}{cc}
M^2_{\squarki{LL}}&M^2_{\squarki{LR}}e^{-i\varphi_{\squark}}\\
M^2_{\squarki{LR}} e^{i\varphi_{\squark}}& M^2_{\squarki{RR}}
\end{array}\right),\quad \tilde{q}=\tilde{U}, \tilde{D}, 
\ee
in the $(\squarki{L},\squarki{R})$ basis, and can be 
diagonalized by a unitary matrix $R_{\squark}$ such that 
\be
(M_{\squark}^{\diag})^2=R_{\squark}^{}M_{\squark}^2 R_{\squark}^{\dagger}. 
\ee
For subsequent discussion it is useful to define the $6\times 3$ matrices 
\be\label{def:stt}
(\G^{q_L})_{ai}=(R_{\squark})_{ai},\quad  
(\G^{q_R})_{ai}=(R_{\squark})_{a, i+3}, \quad q=U,D,
\ee
with $U$ and $D$ denoting up- and down-type quarks respectively. 
Working in the so-called `super-CKM' basis 
\cite{susy:fcnc} in which the $3\times 3$ quark mass matrices $M_U$ and $M_D$ 
are real and diagonal, the submatrices in \eq{squark:massmatrix} take the 
form  
\begin{mathletters}\label{squark:up}
\be
M^2_{\tilde{U}_{LL}}=(M_{\tilde{U}}^2)_{LL} + M_U^2 + 
\frac{1}{6}M_Z^2\cos2\b(3-4\sin^2\theta_W)\openone,
\ee
\be
M^2_{\tilde{U}_{LR}}= M_U |A_U-\m^*\cot\b\openone|,
\ee
\be
M^2_{\tilde{U}_{RR}}=(M_{\tilde{U}}^2)_{RR} + M_U^2 + 
\frac{2}{3}M_Z^2\cos 2\b\sin^2\theta_W\openone,
\ee
\be\label{def:varphiU}
\varphi_{\tilde{U}}= \arg(A_U-\m^*\cot\b\openone),
\ee
\end{mathletters}

\begin{mathletters}\label{squark:down}
\be\label{squark:down:relation}
M^2_{\tilde{D}_{LL}}= \vckm^\dagger (M_{\tilde{U}}^2)_{LL}\vckm + M_D^2 - 
\frac{1}{6}M_Z^2\cos2\b(3-2\sin^2\theta_W)\openone,
\ee
\be
M^2_{\tilde{D}_{LR}}= M_D |A_D-\m^*\tan\b\openone|,
\ee
\be
M^2_{\tilde{D}_{RR}}=(M_{\tilde{D}}^2)_{RR} + M_D^2 - 
\frac{1}{3}M_Z^2\cos 2\b\sin^2\theta_W\openone,
\ee
\be
\varphi_{\tilde{D}}= \arg(A_D-\m^*\tan\b\openone).
\ee
\end{mathletters}%
Here $\theta_W$ denotes the Weinberg angle, 
$\openone$ represents a $3\times 3$ unit matrix, 
$(M_{\tilde{q}}^2)_{LL}$ and $(M_{\tilde{q}}^2)_{RR}$ are  
Hermitian scalar soft mass matrices, and $\vckm$ is the usual CKM matrix. 
In deriving \eq{squark:down:relation}, we have used the relation $(M_{\tilde{D}}^2)_{LL}=
\vckm^\dagger (M_{\tilde{U}}^2)_{LL}\vckm$, which is due to SU(2) gauge invariance. 
Since we ignore flavour-mixing effects among squarks, 
$(M_{\tilde{q}}^2)_{LL}$ and $(M_{\tilde{q}}^2)_{RR}$ in 
Eqs.~(\ref{squark:up}) and (\ref{squark:down}) are diagonal 
-- and hence real -- whereas the $A_q$'s are given by 
\be\label{aterms:diag}
A_U=\diag(A_u, A_c, A_t), \quad A_D=\diag(A_d, A_s, A_b), \quad A_i\equiv 
|A_i|e^{i\varphi_{A_i}}. 
\ee
Consequently, the squark mass-squared matrix, \eq{squark:massmatrix}, 
in the up-squark sector decomposes into a series of $2\times 2 $ matrices. 
As far as the scalar top quark is concerned, we have
\be
M^2_{\tsquark}=
\left(\begin{array}{cc}
m_{\tsquark_L}^2 + m_t^2 + \frac{1}{6}M_Z^2\cos2\b(3-4\sin^2\theta_W) 
&m_t|A_t-\m^*\cot\b|e^{-i\varphi_{\tsquark}}\\
m_t|A_t-\m^*\cot\b|e^{i\varphi_{\tsquark}}& m_{\tsquark_R}^2 + m_t^2 + 
\frac{2}{3}M_Z^2\cos 2\b\sin^2\theta_W\end{array}\right),
\ee
where $m_{\tsquark_L}^2$ and $m_{\tsquark_R}^2$ are diagonal elements of 
$(M_{\tilde{U}}^2)_{LL}$ and $(M_{\tilde{U}}^2)_{RR}$ respectively, while 
$\varphi_{\tsquark}$ can be readily inferred from \eq{def:varphiU}. 
Diagonalization of the stop mass-squared matrix then leads to the physical 
mass eigenstates $\tsquark_1$ and $\tsquark_2 $, namely
\be
\left(\begin{array}{c}
\tsquark_1\\ \tsquark_2\end{array}\right)=
 \left(\begin{array}{cc}
\cos\theta_{\tsquark}& \sin\theta_{\tsquark}e^{-i\varphi_{\tsquark}}\\
-\sin\theta_{\tsquark}e^{i\varphi_{\tsquark}}&\cos\theta_{\tsquark}\end{array}\right)
\left(\begin{array}{c}
\tsquark_L\\ \tsquark_R\end{array}\right)
\equiv 
\left(\begin{array}{cc}
\G^{U_L}_{33}&\G^{U_R}_{33}\\
\G^{U_L}_{63}&\G^{U_R}_{63}\end{array}\right)
\left(\begin{array}{c}
\tsquark_L\\ \tsquark_R\end{array}\right),
\ee
where the mixing angle $\theta_{\tsquark}$ is given by the expression
($-\p/2\leqslant\theta_{\tsquark}\leqslant\p/2$)
\be
\tan 2\theta_{\tsquark}=\frac{2m_t |A_t-\m^*\cot\b|}{(m_{\tsquark_L}^2 -m_{\tsquark_R}^2)+ \frac{1}{6}M_Z^2\cos 2\b(3-8\sin^2\theta_W)}.
\ee

\subsubsection{Chargino mass matrix}
The chargino mass matrix can be written as
\be\label{chargino:mass-matrix}
M_{\chargino}=
\left(
\begin{array}{cc}
M_2 & \sqrt{2} M_W \sin\b\\ \sqrt{2}M_W \cos\b&|\m|e^{i\varphi_{\m}}
\end{array}
\right),
\ee
where we have adopted a phase convention in which the mass term of the 
W-ino field, $M_2$, is real and positive. Note that without loss of 
generality, we can always perform a global rotation to remove one of the three
phases from the gaugino mass parameters $M_i$ ($i=1, 2, 3$).

The mass matrix can be cast in diagonal form by means of a biunitary 
transformation, namely
\be
M_{\chargino}^{\diag}= U^* M_{\chargino}V^{\dagger},
\ee
where $M_{\chargino}^{\diag}$ is diagonal with positive eigenvalues, and 
$U$, $V$ are unitary matrices. Solving the eigenvalue problem
\be\label{chargino:sq}
(M_{\chargino}^{\diag})^2= U^* M_{\chargino}^{}M_{\chargino}^{\dagger}U^T= 
V M_{\chargino}^{\dagger}M_{\chargino}^{}V^{\dagger},
\ee
we find
\be
U=
\left(
\begin{array}{cc}
\cos\theta_U & \sin\theta_U e^{-i\varphi_U}\\ 
-\sin\theta_U e^{i\varphi_U}&\cos\theta_U 
\end{array}
\right),
\ee
\be
V=
\left(
\begin{array}{cc}
\cos\theta_Ve^{-i\phi_1} & \sin\theta_V e^{-i(\varphi_V+\phi_1)}\\ 
-\sin\theta_V e^{i(\varphi_V-\phi_2)}&\cos\theta_V e^{-i\phi_2}
\end{array}
\right),
\ee
with the mixing angles
\be
\tan2\theta_U= \frac{2\sqrt{2}M_W\big[M_2^2\cos^2\beta+|\m|^2\sin^2\beta+|\m|
M_2\sin2\beta\cos\varphi_{\m}]^{1/2}}{M_2^2- |\m|^2-2M_W^2 \cos2\beta},
\ee
\be
\tan2\theta_V= \frac{2\sqrt{2}M_W\big[M_2^2\sin^2\beta+|\m|^2\cos^2\beta+|\m|
M_2\sin2\beta\cos\varphi_{\m}]^{1/2}}{M_2^2- |\m|^2+2M_W^2 \cos2\beta},
\ee
\be
\tan\varphi_U=-\frac{|\m|\sin\varphi_{\m}\sin\beta}{M_2\cos\beta+ 
|\m|\sin\beta\cos\varphi_{\m}},
\ee
\be
\tan\varphi_V=-\frac{|\m|\sin\varphi_{\m} \cos\beta}{M_2\sin\beta+ 
|\m|\cos\beta\cos\varphi_{\m}},  
\ee
\be
\tan\phi_1=\frac{ |\m|\sin\varphi_{\m}M_W^2 \sin2\b}{M_2(m^2_{\tilde{\chi}^{\pm}_1}-|\m|^2)+|\m|M_W^2 \sin2\b\cos
\varphi_{\m}} ,  
\ee
\be
\tan\phi_2=-\frac{ |\m|\sin\varphi_{\m}(m^2_{\tilde{\chi}^{\pm}_2}-M_2^2)}
{M_2M_W^2 \sin2\b+ |\m|(m^2_{\tilde{\chi}^{\pm}_2}-M_2^2)
\cos\varphi_{\m}}.
\ee
Here we have chosen $-\p/2\leqslant\theta_i\leqslant\pi/2$, 
$-\p\leqslant\varphi_i,\phi_i\leqslant\pi$, where $i=U,V$, 
and the chargino mass eigenvalues read 
\bea\label{chargino:mass}
\lefteqn{m^2_{\tilde{\chi}^{\pm}_{1,2}}=\frac{1}{2}\Bigg[M_2^2 + 
|\m|^2 + 2M_W^2}\nnu\\
&\mp& 
\{(M_2^2-|\m|^2)^2+ 4M_W^4\cos^2 2\beta + 4M_W^2 [M_2^2+|\m|^2 + 2 |\m|M_2
\sin2\beta\cos\varphi_{\m}]\}^{1/2}\Bigg].
\eea
%

\subsection{SUSY particles and FCNC interactions}
We present here the SUSY Lagrangian relevant to the FCNC processes of 
interest which will also serve as a means of fixing our 
notation. The interactions of 
charged Higgs bosons, charginos, neutralinos, and gluinos in the presence of
new $\cp$ phases can be written 
as \cite{cho:etal,mssmrev:romao}
\bea
{\cal{L}_{\text{SUSY}}}
&=&\frac{g}{\sqrt{2}M_W}[\cot\b(\bar{u}M_U \vckm P_L d)
+\tan\b(\bar{u}\vckm M_D P_R d)]H^+\nnu\\ 
&+&\sum_{j=1}^2\ol{\tilde{\chi}^-_j}[\tilde{u}^\dagger
(X_j^{U_L}P_L+X_j^{U_R}P_R) d +\tilde{\n}^\dagger
(X_j^{L_L}P_L+X_j^{L_R}P_R)l]\nnu\\
&+&\sum_{k=1}^4\ol{\neutralinoi{k}}[
\tilde{d}^\dagger(Z_k^{D_L}P_L+Z_k^{D_R}P_R) d
+\tilde{l}^\dagger(Z_k^{L_L}P_L+Z_k^{L_R}P_R)l] \nnu\\
&-&\sqrt{2} g_s\sum_{b=1}^8\ol{\tilde{g}^b}
\tilde{d}^\dagger(G^{D_L}P_L-G^{D_R}P_R)T^b d + \hc,
\eea
where generation indices have been suppressed, and $P_{L,R}= (1\mp \g_5)/2$.
The mixing matrices in the super-CKM basis are given by
\begin{mathletters}\label{feyn:X}
\be
X^{U_L}_j=g\Bigg[
{-}V_{j1}^* \G^{U_L} + V_{j2}^* \G^{U_R}
\frac{M_U}{\sqrt{2}M_W \sin\b} \Bigg]\vckm,
\ee
\be
X^{U_R}_j= g U_{j2}\G^{U_L}\vckm\frac{M_D}{\sqrt{2}M_W \cos\b},
\ee
\be
X^{L_L}_j=-g V_{j1}^* R_{\sneutrino}, \quad 
X^{L_R}_j= g U_{j2}R_{\sneutrino}\frac{M_E}{\sqrt{2}M_W \cos\b},
\ee
\be
Z_k^{D_L} =-\frac{g}{\sqrt{2}}\Bigg[\Bigg({-}N_{k2}^*+\frac{1}{3}\tanw 
N_{k1}^*\Bigg)\G^{D_L}+ N_{k3}^*\G^{D_R}\frac{M_D}{M_W \cos\b}\Bigg], 
\ee
\be
Z_k^{D_R} =-\frac{g}{\sqrt{2}}\Bigg[\frac{2}{3}\tanw N_{k1}\G^{D_R} + 
N_{k3}\G^{D_L}\frac{M_D}{M_W \cos\b}\Bigg],
\ee
\be
Z_k^{L_L} =\frac{g}{\sqrt{2}}\Bigg[(N_{k2}^*+\tanw N_{k1}^*)\G^{L_L}- 
N_{k3}^*\G^{L_R}\frac{M_E}{M_W \cos\b}\Bigg], 
\ee
\be
Z_k^{L_R} =-\frac{g}{\sqrt{2}}\Bigg[2\tanw N_{k1}\G^{L_R} + 
N_{k3}\G^{L_L}\frac{M_E}{M_W \cos\b}\Bigg],
\ee
\be
G^{D_L}=e^{-i\varphi_3/2}\G^{D_L},\quad 
G^{D_R}=e^{i\varphi_3/2}\G^{D_R},
\ee
\end{mathletters}%
$\varphi_3$ being the phase of the 
gluino mass term $M_3$. (For scalar lepton as well as neutralino mass 
and mixing matrices, we refer the reader to \rf{mssmrev:romao}.)  
In the remainder of this section, we briefly discuss two SUSY 
models with quite distinct scenarios for the $\cp$-violating 
phases. 

\subsection{Different scenarios for the SUSY parameters}
\subsubsection{Constrained MSSM}
In order to  solve the FCNC problem in the MSSM and to 
further reduce the number of unknown parameters, the MSSM is 
generally embedded in a grand unified theory (GUT). 
This leads to the minimal supergravity (mSUGRA) inspired model, commonly  
referred to as the constrained MSSM \cite{mssm:original}. In this model one 
assumes universality of the soft terms at some high scale, which we take to 
be the scale of gauge coupling unification, $\mgut$, 
implying that (i) all gaugino mass parameters are equal to a common 
mass $M_{1/2}$; (ii) all the scalar mass parameters share a common value 
$m_0$; and (iii) all the soft trilinear couplings are equal to $A_0$. 
As a result, the mSUGRA model has only two new independent phases 
which are associated with the $\m_0$ and $A_0$ parameters. 
After all, we have at the GUT scale 
\be\label{set:modelI:GUT}
\tan\beta,\ M_{1/2},\ m_0,\ |A_0|,\ |\m_0|,\ \varphi_{A_0},\ \varphi_{\m_0},
\ee 
with $M_{1/2}$ and $m_0$ being real. The parameters at the electroweak 
scale are then obtained by solving the renormalization group equations 
(RGE's).

A few remarks are in order here. 
First, the phases of the gaugino mass terms 
$M_i$ are not affected by the renormalization group evolution, and therefore 
the low energy gaugino mass parameters are real. 
Second, the phase that appears together with the 
$\m$ parameter does not run at one-loop level
so that $\varphi_{\m}=\varphi_{\m_0}$. Moreover, to satisfy the 
constraints on the 
EDM's of elec\-tron and neutron, $\varphi_{\m}$ has to be of 
${\mathcal O}(10^{-2})$ unless strong cancellations between different 
contributions occur.~Third, solving the RGE's for the evolution of the $\cp$-violating phase 
of the $A_t$-term yields a small value for $\varphi_{A_t}$, 
and thus is not constrained by the EDM's 
\cite{goto:etal:phase:sugra,phase:sugra}. 
Lastly, off-diagonal entries occur in the squark mass matrices due to 
renormalization group evolution of the parameters even in the absence of 
flavour mixing at the GUT scale. However, these effects are found to be 
small and therefore the squark mass matrix is essentially flavour 
diagonal at the electroweak scale (see also \rfs{cho:etal,offdiag:cmssm}).

\subsubsection{Effective supersymmetry} 
As an example of SUSY models with large CP phases, 
we consider the effective supersymmetry picture 
\cite{effsusy} without assuming universality of sfermion masses at a high 
scale. Within such a framework, the first and second generation sfermions 
are almost degenerate and have masses above the $\TeV$ scale, 
while third generation sfermions can be light enough to be 
accessible at future hadron colliders. Consequently, FCNC reactions 
as well as one-loop contributions to the EDM's of electron and neutron 
are well below the current experimental bounds.

However, it should be noted that the EDM's also receive contributions at 
two-loop level involving scalar bottom and top quarks that may become 
important for phases of order unity in the large 
$\tan\beta$ regime \cite{edm:two-loop}. 
In our numerical work, $\tan\b$ is assumed to be in the interval 
$2\leqslant \tan\b\leqslant 30$.

\section{Rare \bm $B$ decays and new physics}\label{rareBdecays}
\subsection{Short-distance matrix element}
Let us start with the QCD-corrected matrix element describing the 
short-distance interactions in $b\to s(d) l^+l^-$ within the SM. 
It is given by 
\bea\label{mael:short}
\lefteqn{{\cal M}_{\sd}= \frac{G_F \a}{\sqrt{2} \p}V_{tb}^{}V_{tf}^{\ast}
\Bigg\{\Bigg[(\ceff-c_{10}) \braket{H(k)}{\bar{f}\g_{\m}P_L b}{\B(p)}}\nnu\\
& & \mbox{}- \frac{2\cseff}{q^2}\braket{H(k)}{\bar{f}i \s_{\m\n}q^{\n}\left(m_b P_R+ m_f P_L\right) b}{\B(p)}\Bigg]\bar{l}\g^{\m} P_L l + 
(c_{10}\to -c_{10}) \bar{l}\g^{\m} P_R l\Bigg\}, 
\eea
where $q$ is the four-momentum of the lepton pair, 
and $H=K, K^*$ ($\p,\r$) in the case of $f=s$ ($f=d$).
In the SM, the Wilson coefficients $\cseff$ and $c_{10}$ are real with  
values of $-0.314$ and $-4.582$ respectively, and the leading term in $\ceff$ has the form \cite{bmm,review}
\be\label{wilson:c9app}
\ceff=c_9 +  (3 c_1 + c_ 2)\Big\{g(m_c,q^2) + \l_u\Big[g(m_c,q^2)
-g(m_u,q^2)\Big]\Big\}+ \cdots ,
\ee 
where $c_9=4.216$. The Wilson coefficients will be discussed in detail in Appendix 
\ref{app:wilson}. In the above expression 
\be\label{lambdau}
\l_u \equiv \frac{V_{ub}^{}V_{uf}^*}{V_{tb}^{}V_{tf}^*}\approx 
\left\{\begin{array}{l}-\l^2(\r-i\eta) \quad \text{for}\quad f=s ,\\
\dis\frac{\rho(1-\rho)-\eta^2}{(1-\rho)^2 + \eta^2} 
- \frac{i \eta}{(1-\rho)^2 + \eta^2} \quad \text{for}\quad f=d,
\end{array}\right.
\ee
with $\rho$ and $\eta$ being the Wolfenstein parameters 
\cite{ckm:wolfenstein}, where the latter reflects the presence of $\cp$ violation in the SM. For definiteness, we will assume $\rho=0.19$ and 
$\eta=0.35$ \cite{ckm:analysis}. 

Finally, the one-loop function $g(m_i,q^2)$ at the scale $\m_R=m_b$ 
is given by\footnote{In order to avoid confusion with the $\m$ parameter of 
the superpotential, we use the notation $\m_R$ for the renormalization scale.}
\bea\label{loopfunc}
\lefteqn{g(m_i,q^2)=-\frac{8}{9}\ln(m_i/m_b)+\frac{8}{27}+\frac{4}{9}y_i
-\frac{2}{9}(2+y_i)\sqrt{|1-y_i|}}\nnu\\[.7ex]
&&\times\left\{
\Theta(1-y_i)\left[\ln\left(\frac{1 + \sqrt{1-y_i}}{1 - \sqrt{1-y_i}}\right)-i\p\right]+ \Theta(y_i-1) 2\arctan\frac{1}{\sqrt{y_i-1}}\right\},
\eea
where  $y_i = 4m_i^2/q^2$. Observe that $c\bar{c}$ and $u\bar{u}$ 
loops provide absorptive parts that are mandatory, as we show
below, for a non-zero partial width asymmetry besides the presence of a $\cp$-violating phase.

\subsection{Wilson coefficients and new physics} 
Throughout this paper, we will assume that in the presence of new 
physics there are no new operators beyond those that correspond to the 
Wilson coefficients appearing in \eq{mael:short}.
(For a discussion of the implications of new operators for rare $B$ 
decays, see, e.g., \rf{ops:new}.) 
Thus, the effect of new physics is simply 
to modify the matching conditions of the Wilson coefficients, i.e.~their 
absolute values and phases at the electroweak scale.

As a result, we are left with additional SUSY contributions at one-loop level 
to the Wilson coefficients $\cseff$, $\ceff$, and $c_{10}$ in \eq{mael:short}. 
In fact, they arise from penguin and box diagrams 
with (i) charged Higgs boson up-type quark loops;
(ii) chargino up-type squark loops; (iii) 
neutralino down-type squark loops; and (iv) gluino down-type squark loops.
Thus, the short-distance coefficients can be conveniently written as
\be\label{wilson:komplett}
c_i(M_W) = c_i^{\sm}(M_W) + c_i^{\higgs}(M_W) + c_i^{\chargino}(M_W)
 + c_i^{\neutralino}(M_W)+ c_i^{\gluino}(M_W) \quad (i=7, \dots, 10).
\ee
The explicit expressions for the various Wilson coefficients 
are given in Appendix \ref{app:wilson}.
Since we limit our attention to flavour-conserving effects in the 
squark sector, the neutralino and gluino exchange contributions in 
\eq{wilson:komplett} will be omitted in our numerical calculations. 

For future reference, we parametrize the new physics contributions as follows:
\be\label{ratio:np-sm}
R_i = \frac{c_i(M_W)}{c_i^{\sm}(M_W)}\equiv |R_i| e^{i \phi_i},
\quad \chi=\frac{R_8-1}{R_7-1},
\ee
where $\chi$ is real to a good approximation within the models under study. 

\subsection{Resonant intermediate states}
We have considered so far only the short-distance interactions.
A more complete ana\-ly\-sis, however, has also to take into account 
resonance contributions due to $u\bar{u}$, $d\bar{d}$, and $c\bar{c}$ 
intermediate states, i.e.~$\r, \omega, J/\psi, \psi'$, and so forth. 
A detailed discussion of the various theoretical suggestions of how to  
describe these effects is given in \rf{ali:hiller}. 

We employ here the approach proposed in \rf{fklms:res} 
which makes use of the renormalized photon vacuum polarization,
$\Pi^\g_\hadron (s)$, related to cross-section 
data\footnote{This
method assumes quark-hadron duality and rests on the factorization 
assumption.}
\be
\rhadron(s)\equiv \frac{\s_{\text{tot}}(e^+e^-\to {\text{hadrons}})}{\s(e^+e^-\to \m^+\m^-)},
\ee
with $s\equiv q^2$.
In fact, the absorptive part of the vacuum polarization is given by  
\be
\Im \Pi^\g_\hadron (s) = \frac{\a}{3}\rhadron(s), 
\ee
whereas the dispersive part may be obtained via a once-subtracted dispersion 
relation \cite{dispersive}
\be
\Re  \Pi^\g_\hadron (s)= \frac{\a s}{3\p}P \int\limits_{4M_{\p}^2}^\infty\frac{\rhadron(s')}{s'(s'-s)}d s', 
\ee
with $P$ denoting the principal value.~For example, in the case of the 
$J/\psi$ family (i.e.~$J/\psi, \psi', \dots$) the imaginary part of
the one-loop function $g(m_c,s)$, \eq{loopfunc}, can be expressed as 
\be\label{im:gmc}
\Im g(m_c, s)=\frac{\p}{3}\rhadron^{J/\psi}(s),\quad 
\rhadron^{J/\psi}(s)\equiv \rcontcc (s)+ \rrespsi (s),  
\ee
where the subscripts `cont' and `res' stand for continuum and resonance 
contributions respectively, while the real part is given by
\be\label{re:gmc}
\Re  g(m_c, s)= -\frac{8}{9}\ln(m_c/m_b)-\frac{4}{9}+ \frac{s}{3}
P \int\limits_{4M_{\p}^2}^\infty\frac{\rhadron^{J/\psi}(s')}{s'(s'-s)}d s'. 
\ee
The contributions from the continuum can be determined by means of 
experimental data given in \rf{burkhardt}, whereas the narrow resonances 
are well described by a relativistic Breit-Wigner distribution.

However, in order to reproduce correctly the branching ratio 
for direct $J/\psi$ production via the relation 
$(V_{c\bar{c}}=J/\psi, \psi',\dots$)
\be\label{rel:factorize}
\branch(B\to H V_{c\bar{c}}\to H l^+l^-)=
\branch(B\to H V_{c\bar{c}})\branch(V_{c\bar{c}}\to l^+l^-), 
\ee 
where $H$ stands for pseudoscalar and scalar mesons, 
one has to multiply $\rrespsi$ in 
Eqs.~(\ref{im:gmc}) and (\ref{re:gmc}) by a phenomenological 
factor $\k$, regardless of which method one uses for the description of the 
resonances  \cite{amm,lw}.\footnote{Strictly speaking, it is 
the term $(3c_1+c_2)\rrespsi$ -- in the approximation of \eq{wilson:c9app} -- 
which has to be corrected to $\k(3c_1+c_2)\rrespsi$, taking into account  
non-factorizable contributions in two-body $B$ decays (see, e.g., 
\rf{non-fac:review}).}
Using the form factors of \rf{melikhov} (see next section)
together with experimental data on 
$\branch(B\to K^{(*)}J/\psi)$, $\branch(B\to K^{(*)}\psi')$, and 
$\branch(B^-\to \p^- J/\psi)$ \cite{pdg}, 
we find a magnitude for $\k$ of $1.7$ to $3.3$. At this point two 
remarks are in order. First, the branching ratio for direct 
$J/\psi$ and $\psi'$ production is 
enhanced by a factor $\k^2$, while it is essentially independent of 
$\k$ outside the resonance region.
Second, the numerical results for average $\cp$ asymmetries
in the non-resonant continuum $1.2\, \GeV < \sqrt{s} < 2.9\, \GeV$
are not affected by the uncertainty in $\k$.

Similar considerations also hold for $u\bar{u}$ and $d\bar{d}$ systems 
except that the $\r$ resonance is described through
\be
\rresrho(s) = \frac{1}{4}\left(1-\frac{4M_{\p}^2}{s}\right)^{3/2}|F_{\p}(s)|^2,
\ee
where the pion form factor is given by a modified Gounaris-Sakurai 
formula \cite{gounsak}. 
 
\section{The decays \bm$\B\to K^{(*)}\, \lowercase{l^+l^-}$
and \bm$\B\to \p (\r)\, \lowercase{l^+l^-}$}\label{decays:exc}
The hadronic matrix elements in exclusive $B$ decays can be written 
in terms of $q^2$-dependent form factors, where $q^2$ is the 
invariant mass of the lepton pair. 
In the work described here, we employ heavy-to-light 
$B\to K^{(*)}$ and $B\to \p (\r)$ form factors determined 
by Melikhov and Nikitin \cite{melikhov}
within a relativistic quark model. To get an estimate of the 
theoretical uncertainty that is inherent to any model for the form 
factors, we also utilize the parametrization of 
Colangelo \ea\ \cite{colangelo}, which makes use of QCD sum rule 
predictions.

For simplicity of presentation, we do not display corrections due to a 
non-zero lepton mass, which can be found in 
\rfs{fklms:exc,melikhov:formulae}. (The same remark 
applies to the light quark masses $m_{s,d}$.) Henceforth 
we shall denote pseudoscalar and vector mesons
by $P=K, \p$ and $V=K^*, \r$ respectively.

\subsection{\bm$B\to P$ transitions}
\subsubsection{Form factors}
The hadronic matrix elements for the decays $B\to P$ can be 
parametrized in terms of three Lorentz-invariant form factors 
(see Appendix \ref{formfactors} for details), namely 
\begin{mathletters}
\be
\braket{P(k)}{\bar{f}\g_{\m}P_L b}{\B(p)}=\frac{1}{2}[
(2p-q)_{\m} f_+(q^2)+ q_{\m} f_-(q^2)],
\ee
\be
\braket{P(k)}{\bar{f}i\s_{\m\n}q^{\n}P_{L,R}b}{\B(p)}=
-\frac{1}{2}[(2p - q)_{\m}q^2 - (M_B^2 - M_P^2)q_{\m}]s(q^2),
\ee
\end{mathletters}%
with $q=p-k$. Here we assume that the form factors are real, 
in the absence of final-state interactions. 
Note that the terms proportional to $q_{\m}$ may be 
dropped in the case of massless leptons. 

\subsubsection{Differential decay spectrum and forward-backward asymmetry}
Introducing the shorthand notation 
\be\label{def:triangle}
\l (a,b,c) = a^2 + b^2 + c^2 - 2 (a b + b c + a c),
\ee
\be\label{def:X}
X_i =\frac{1}{2}\l^{1/2}(M_B^2, M_i^2, s),
\ee
and recalling $s\equiv q^2$, the differential decay rate can be written 
as (neglecting $m_l$ and $m_f$)
\be\label{diff:BtoP}
\frac{d\G(\B\to P l^+l^-)}{d s\, d\cos\theta_l}
=\frac{G_F^2 \a^2}{2^8 \p^5 M_B^3}\,|V_{tb}^{}V_{tf}^{\ast}|^2 X_P^3
\bigg[|\ceff f_+(s) + 2\cseff m_b s(s)|^2 + |c_{10}f_+(s)|^2\bigg]\sin^2\theta_l.
\ee
Here $\theta_l$ is the angle between $l^-$ and the outgoing hadron
in the dilepton centre-of-mass system, and the Wilson coefficients are 
collected in Appendix \ref{app:wilson}. Defining the forward-backward (FB) 
asymmetry as
\be\label{FB}  
A_{\text{FB}}(s)=\frac{\dis\int_0^1 d\cos\theta_l
\frac{d\G}{ds\, d\cos\theta_l}-\int_{-1}^0 d\cos\theta_l
\frac{d\G}{ds\, d\cos\theta_l}}{\dis\int_0^1 d\cos\theta_l
\frac{d\G}{ds\, d\cos\theta_l}+\int_{-1}^0 d\cos\theta_l
\frac{d\G}{ds\, d\cos\theta_l}},
\ee
which is equivalent to the energy asymmetry discussed in \rf{cho:etal},
it follows directly from the distribution in \eq{diff:BtoP} 
that $A_{\text{FB}}$ vanishes in the case of $\B\to Pl^+l^-$ transitions.
We note in passing that, given an extended operator basis (e.g.~in models 
with neutral Higgs-boson exchange), new Dirac 
structures $\bar{l}l$ and $\bar{l}\g_5l$ may occur in \eq{mael:short}, 
giving rise to a non-zero FB asymmetry in $\B\to Pl^+l^-$.

\subsection{\bm $B\to V$ transitions}
\subsubsection{Form factors}
The hadronic matrix elements describing the decays $B\to V$ 
are characterized by seven independent form 
factors, which we present in Appendix \ref{formfactors}, 
defined through ($\epsi_{0123} = +1$)
\begin{mathletters}\label{def:formfactors:btov}
\bea
\lefteqn{\braket{V(k)}{\bar{f}\g_{\m}P_L b}{\bar{B}(p)}}\nnu\\
&&=i
\epsi_{\m\n\a\b}\epsi^{\n*}p^{\a}q^{\b} g(q^2) - \frac{1}{2}\Bigg\{
\epsi_{\m}^* f(q^2) + (\epsi^*\cdot q)[(2p -q)_{\m}a_+(q^2)+ q_{\m}a_-(q^2)]\Bigg\},
\eea 
\bea
\lefteqn{\braket{V(k)}{\bar{f}i \s_{\m\n}q^{\n}P_{R,L} b}{\bar{B}(p)}
=i\epsi_{\m\n\a\b}\epsi^{\n*}p^{\a}q^{\b}g_+(q^2)
\mp\frac{1}{2}\epsi_{\m}^*[g_+(q^2)(M_B^2-M_V^2)+ q^2 g_-(q^2)]}
\nnu\\
&\pm& \frac{1}{2}(\epsi^*\cdot q)\Bigg\{(2p-q)_{\m}\Bigg[g_+(q^2) + 
\frac{1}{2}q^2h(q^2)\Bigg]+ q_{\m}\Bigg[g_-(q^2) -\frac{1}{2}(M_B^2-M_V^2)h(q^2)\Bigg]\Bigg\},
\eea
\end{mathletters}%
$\epsi^{\m}$ being the polarization vector of the final-state meson, 
and $q=p-k$.

\subsubsection{Differential decay spectrum and forward-backward asymmetry}
The differential decay rate for $\B\to Vl^+l^-$ in the case of massless 
leptons and light quarks takes the form ($f=s$ or $d$) 
\be\label{diff:BtoV}
\frac{d\G(\B\to V l^+l^-)}{d s\, d\cos\theta_l}
=\frac{G_F^2 \a^2}{2^9 \p^5 M_B^3}\,|V_{tb}^{}V_{tf}^{\ast}|^2 X_V
\bigg[A(s) + B(s) \cos\theta_l + C(s) \cos^2\theta_l\bigg],
\ee
with $X_V$ as in \eq{def:X}. The quantities $A$, $B$, and $C$ are  
\bea\label{double:A}
A(s) &=& \frac{2X_V^2}{M_V^2}\Bigg[s M_V^2 \a_1(s) + \frac{1}{4}\Bigg(1 + \frac{2 s M_V^2}{X_V^2}\Bigg) \a_2(s)
+ X_V^2 \a_3(s)+ (k\cdot q)\a_4(s)\Bigg],
\eea
\be\label{double:B}
B(s) = 8  X_V \Re\{c_{10}^*[\ceff s A_x A_y - \cseff m_b (A_x B_y + A_y B_x)]\},
\ee
\be\label{double:C}
C(s) =  \frac{2X_V^2}{M_V^2}\Bigg[s M_V^2 \a_1(s) - \frac{1}{4} \a_2(s)- X_V^2 \a_3(s) - (k\cdot q) \a_4(s)\Bigg],
\ee
where $k\cdot q = (M_B^2 - M_V^2-s)/2$, and 
\begin{mathletters}\label{definition:alphai}
\be
\a_1(s ) = (|\ceff|^2 + |c_{10}|^2) A_x^2 + 
\frac{4|\cseff|^2m_b^2}{s^2}B_x^2
-\frac{4\Re(\cseff\ceffstar) m_b}{s} {A}_x {B}_x, 
\ee
\be
\a_2(s) = {\a_1(s )}_{x\to y}, \quad \a_3 (s) = {\a_1(s )}_{x\to z},
\ee
\be
\a_4(s) = (|\ceff|^2 + |c_{10}|^2) {A}_y {A}_z + \frac{4|\cseff|^2m_b^2}{s^2} {B}_y{B}_z
-\frac{2\Re(\cseff\ceffstar)m_b}{s}(A_y B_z+A_z B_y),
\ee
\end{mathletters}%
in which the $A_i$'s and $B_i$'s are defined as 
\begin{mathletters}
\be
A_x=g(s),\quad A_y= f(s),\quad A_z= a_+(s), 
\ee
\be
B_x= g_+(s), \quad B_y= g_+(s)(M_B^2-M_V^2)+ s g_-(s), \quad 
B_z= -\Bigg[g_+(s)+ \frac{1}{2}s h(s)\Bigg].
\ee
\end{mathletters}%
The complex Wilson coefficients $\cseff$, $\ceff$, and $c_{10}$ 
are given in Appendix \ref{app:wilson}.

Finally, using Eqs.~(\ref{FB}) and (\ref{diff:BtoV}), 
we derive the forward-backward asymmetry 
\be\label{fb:BtoV}
A_{\text{FB}}(s)= 12 X_V \frac{\Re\{c_{10}^*[\ceff s A_x A_y - \cseff m_b 
(A_x B_y + A_y B_x)]\}}{[3A(s)+ C(s)]}.
\ee

\subsection{CP-violating observables}
To discuss $\cp$-violating asymmetries, let us first recall
the necessary ingredients. Suppose the decay amplitude  
for $\B\to F$ has the general form  
\be\label{matrixelement:itof}
{\cal A}(\B\to F)=e^{i\phi_1}A_1 e^{i\d_1} +e^{i\phi_2}A_2 e^{i\d_2},   
\ee
where $\d_i$ and $\phi_i$ denote  strong phases ($\cp$-conserving) and 
 weak phases ($\cp$-violating) respectively ($A_1$ and $A_2$ being real).
Together with the decay amplitude for the conjugate process
\be
\ol{\cal A}(B\to \bar{F})
=e^{-i\phi_1}A_1 e^{i\d_1} +e^{-i\phi_2}A_2 e^{i\d_2},
\ee
which can be obtained from \eq{matrixelement:itof} by means of $\cpt$ 
invariance, we may define the $\cp$ asymmetry as
\be\label{def:CPasym:gen}
A_{\cp}\equiv\frac{|{\cal A}|^2-|\ol{\cal A}|^2}{|{\cal A}|^2
+|\ol{\cal A}|^2}= \frac{-2 r\sin\phi\sin\d}{1 + 2r\cos\phi\cos\d+ r^2},
\ee
with $r=A_2/A_1$, $\phi=\phi_1-\phi_2$, and $\d=\d_1-\d_2$. As can be easily seen from 
\eq{def:CPasym:gen}, a non-zero 
partial rate asymmetry requires the simultaneous presence of a 
$\cp$-violating phase $\phi$ as well as a $\cp$-conserving dynamical 
phase $\d$, the latter being provided by the one-loop functions 
$g(m_c,s)$ and $g(m_u,s)$ present in the Wilson coefficient $\ceff$
[\eq{wilson:c9app}]. Notice that in the limit in which the charm quark mass 
equals the up quark mass there is no $\cp$ violation in the SM.

Given the differential decay distribution in the variables $s$ and 
$\cos\theta_l$, we can construct the following $\cp$-violating 
observables:
\begin{mathletters}\label{def:CPasym}
\be\label{def:CPasymSD}
{A_{\cp}^{D,S}}(s)=\frac{\dis \int_{D,S} d\cos\theta_l\frac{d\G_{\text{diff}}}{d s\, d\cos\theta_l}}{\dis \int_S d\cos\theta_l\frac{d\G_{\text{sum}}}{d s\, d\cos\theta_l}}, 
\quad \int_{D, S}\equiv \int_0^{1}\mp\int_{-1}^0,
\ee
where we have introduced
\be
\Gamma_{\text{diff}}=\Gamma(\bar{B}\to H l^+l^-)- 
\ol{\G}(B \to\bar{H}l^+l^-),
\ee
\be
\Gamma_{\text{sum}}=\Gamma (\bar{B}\to H l^+l^-)+ 
\ol{\G}(B \to\bar{H}l^+l^-),
\ee
\end{mathletters}%
with $H=K^{(*)}, \p,\r$. It should be noted that the 
asymmetry $A_{\cp}^D$ represents a $\cp$-violating effect in the 
angular distribution of $l^-$ in $B$ and $\B$ decays 
while $A_{\cp}^S$ is the asymmetry in the partial widths of 
these decays. As can be seen from Eqs.~(\ref{diff:BtoP}), (\ref{diff:BtoV}), 
and (\ref{fb:BtoV}), the latter involves the phases of $\cseff$ and $\ceff$ 
while the former is also sensitive to the phase of the Wilson coefficient 
$c_{10}$.    

\section{Numerical analysis}\label{results}
Given the SUSY contributions presented in the preceding sections, 
we now proceed to study the implications of supersymmetry
for exclusive $B$ decays. 

\subsection{Experimental constraints}
\begin{figure}[ht]
\epsfig{figure=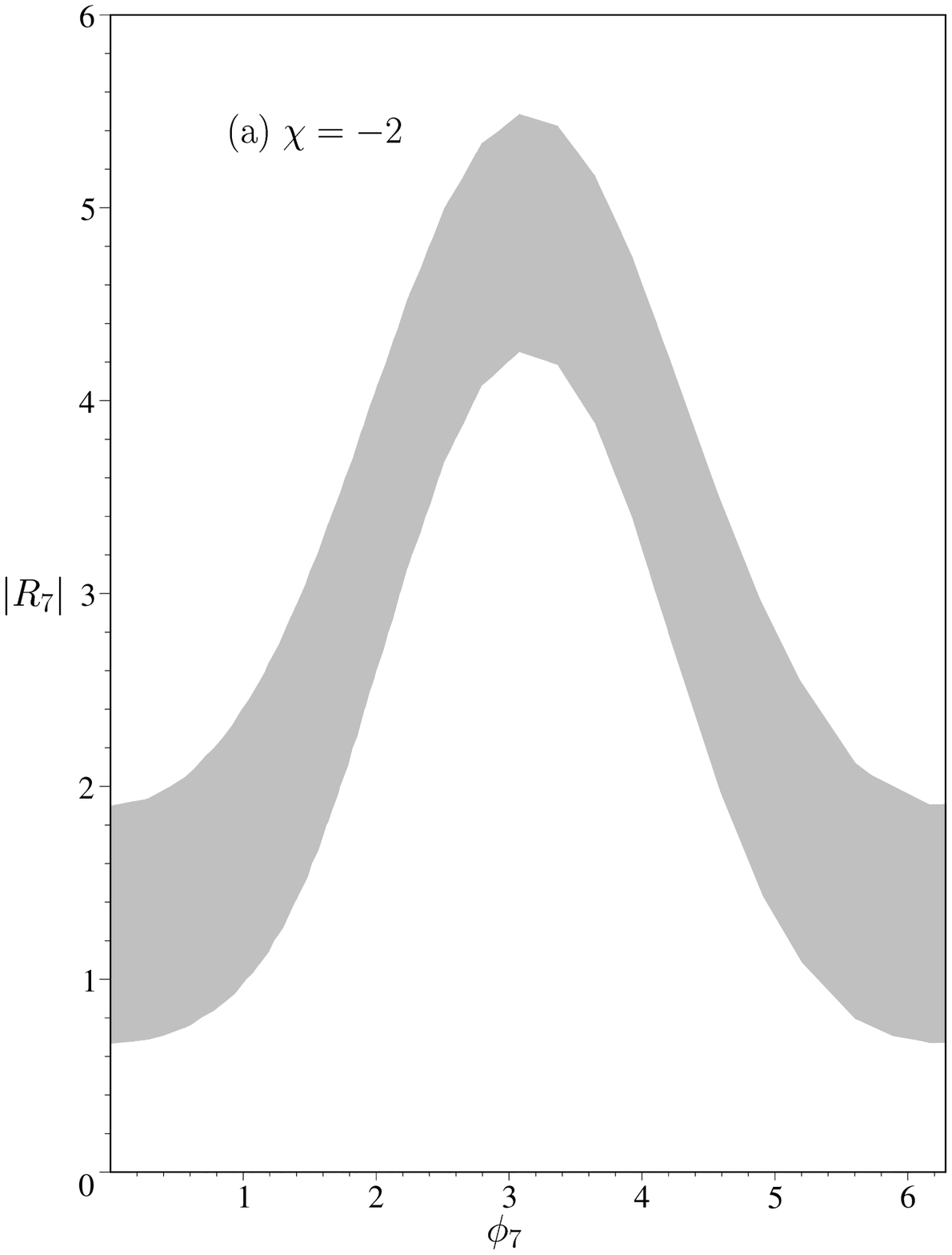,height=4.1in,angle=0}\hspace{1em}
\epsfig{figure=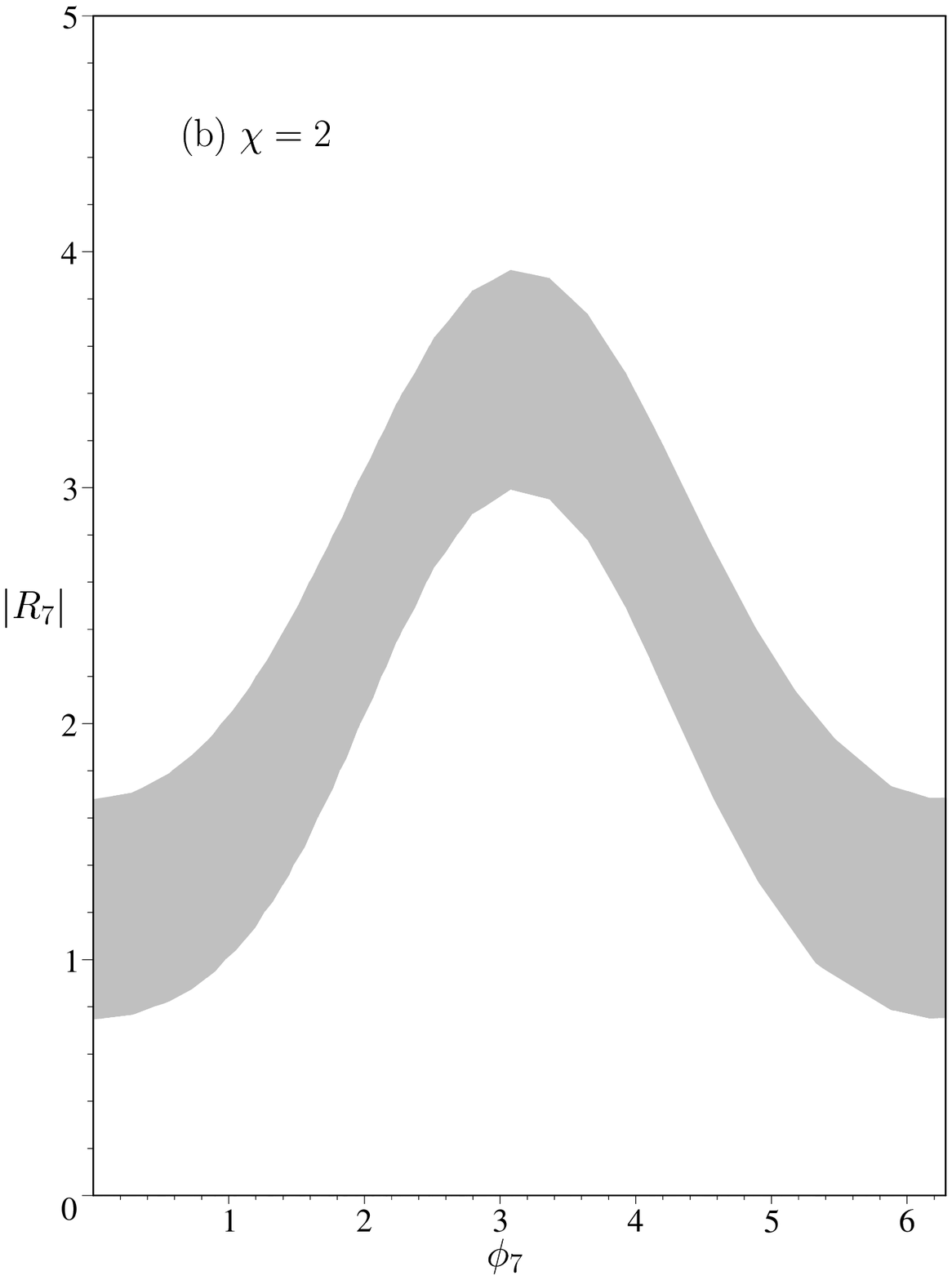,height=4.1in,angle=0}\vspace{2em}
\caption[]{Allowed region for $|R_7|$ and the corresponding 
$\cp$-violating phase $\phi_7$ as determined from the inclusive 
measurement of $b\to s\g$ rate, using the leading-log expression for 
$\cseff$. Diagrams (a) and (b) correspond to different values of $\chi$ 
[\eq{ratio:np-sm}].\label{r7bsg}}
\end{figure}

In our numerical analysis, we scan the SUSY parameter space as given 
in \rf{baek:ko} and take as input the parameters displayed
in Appendix \ref{app:numerics}. In addition, we take into account the 
following experimental constraints: 
\bit
\item From the measurement of the inclusive branching ratio 
$\branch(\B\to X_s \g)$, which probes $|\cseff|$, one can derive upper 
and lower limits \cite{exp:btosg}:
\be\label{range:btosg}
2.0\times 10^{-4}< \branch(\B\to X_s \g)<4.5\times 10^{-4}
\quad (\cl{95}).
\ee
This is specially useful to constrain extensions of the SM. Indeed, 
following the model-independent analysis performed in 
\rf{kagan:neubert:btosg}, and taking the Wilson coefficient 
$\cseff$ in leading-log approximation [see \eq{wilson:c7eff} of the Appendix],
we obtain 
\bea\label{btosg:theo}
\branch(\B\to X_s\g)&\approx& [0.801 + 0.444 |R_7|^2 + 0.002 |R_8|^2 
+ 1.192\Re R_7 \nnu\\&&\mbox{}+ 0.083\Re R_8+ 
0.061 \Re(R_7 R_8^*)]\times 10^{-4}, 
\eea
where $R_7$ and $R_8$ are as in \eq{ratio:np-sm}. From \fig{r7bsg}
we infer that the present $b\to s \g$ measurement already excludes 
many solutions for $R_7$. 
%
%
\item A CDF search for the exclusive decays of interest yields the 
upper limits $\branch(B^0\to K^{*0} \m^+\m^-) < 4.0\times 10^{-6}$ and 
$\branch(B^+\to K^+ \m^+\m^-) <5.2\times 10^{-6}$ at the $\cl{90}$ 
\cite{limits:exc:exp}. Note that the $K^{*0} \m^+\m^-$ upper limit is 
close to the SM prediction \cite{np2:exc,SMprediction}.~As for the modes $B\to \p l^+l^-$ and
$B\to \r l^+l^-$, we are not aware of any such limits.

\item Non-observation of any SUSY signals at LEP 2 and the 
Tevatron imposes the following lower bounds \cite{pdg}:
\bea
m_{\chargino}> 86\, \GeV, m_{\sneutrino} > 43\, \GeV,
m_{\tsquarki{1}}> 86\, \GeV, m_{\squark} > 260\, \GeV,
m_{H^\pm}> 90\, \GeV.
\eea
\eit

\subsection{\bm$\cp$ asymmetries} 
\begin{figure}[ht]
\epsfig{figure=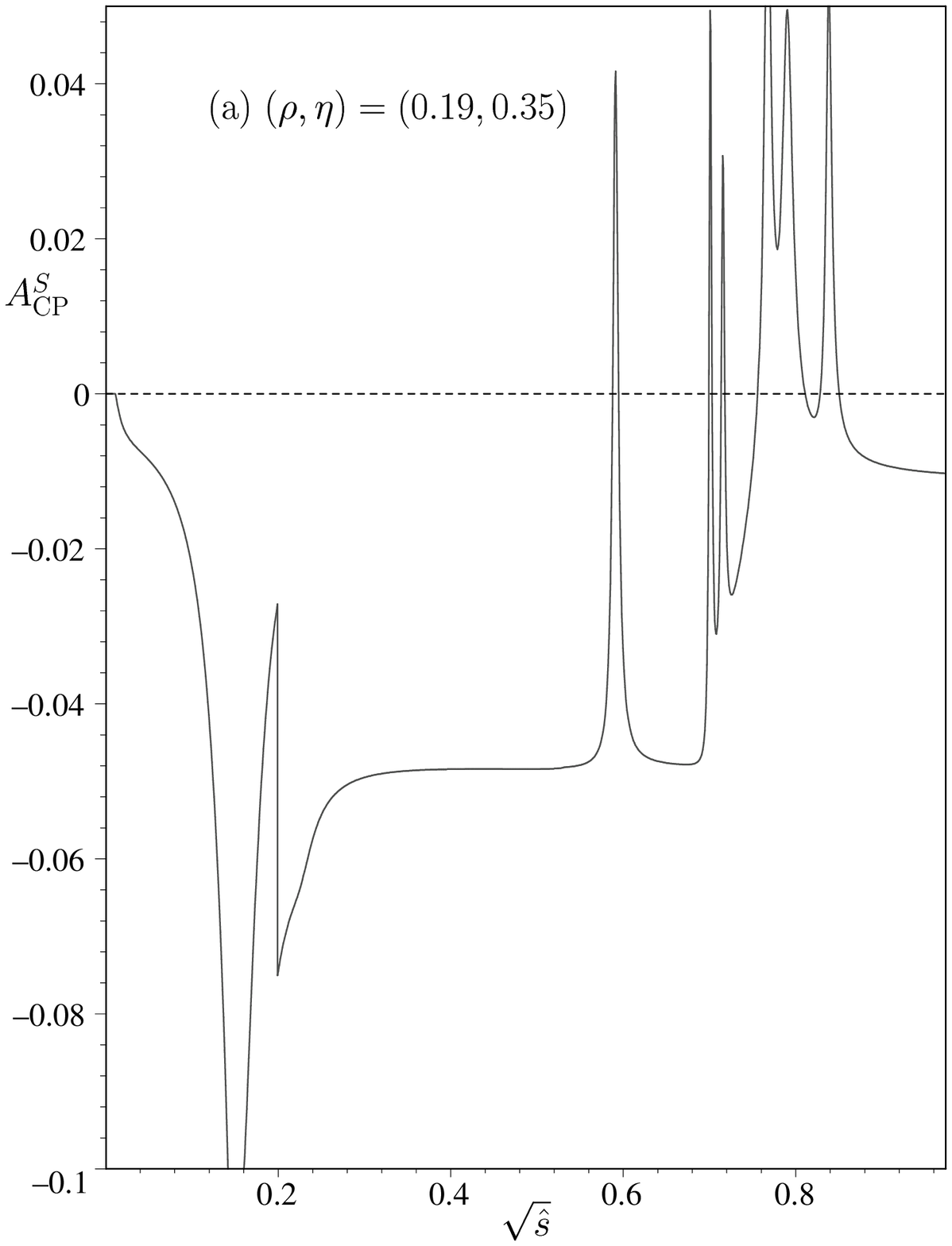,height=4in,angle=0}\hspace{1em}
\epsfig{figure=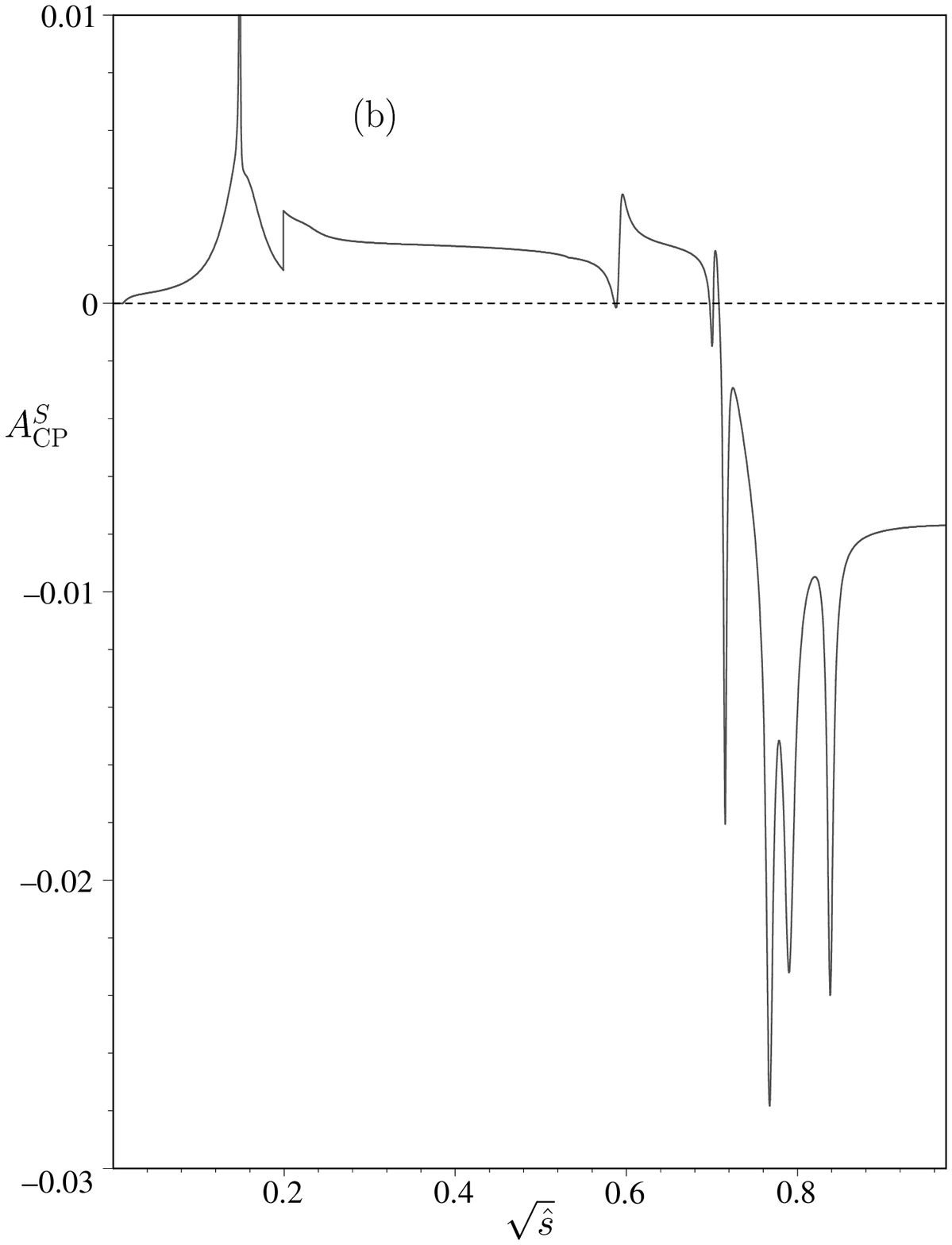,height=4in,angle=0}\vspace{2em}
\caption[]{$\cp$-violating partial width asymmetry $A_{\cp}^S$ in the decays
$B^-\to \p^- l^+l^-$ and $B^+\to \p^+ l^+l^-$ vs $\sqrt{\hat{s}}$, 
$\hat{s}\equiv s/M_B^2$, including $\r$, $\omega$, and $J/\psi$, $\psi'$, 
etc.~resonances, and employing the form factors of \rf{melikhov}. (a) Within 
the SM and (b) in the presence of new $\cp$ phases and a real CKM matrix. For 
the sake of illustration, we have chosen $|R_7|=1.6$, $|R_{9,10}|=1$, 
$\chi=1.5$, $\phi_7=\p/2$, and $\phi_{9,10}=0.01$.\label{acpSM}}
\end{figure}

As mentioned earlier, we investigate $\cp$ asymmetries in the low 
dilepton invariant mass region, i.e.~$1.2\, \GeV < \sqrt{s} < (M_{J/\psi}-200\, \MeV)$, which is of particular interest because the low-$s$ region is 
sensitive to the Wilson coefficient $\cseff$ (in the case of $B\to V l^+l^-$). 
In fact, it can receive large SUSY contributions and be complex,
as well as change sign, while being consistent with 
the experimental measurement of $b\to s \g$. 
On the other hand, the new-physics effects are known to alter the 
remaining Wilson coefficients $\ceff$ and $c_{10}$ only slightly
within mSUGRA and effective SUSY with no additional flavour structure 
beyond the usual CKM mechanism \cite{joanne,goto:etal,np2:exc}. Moreover,
$\cp$ asymmetries above the $J/\psi$ resonance are 
dominated by 
$c\bar{c}$ resonant intermediate states, whereas below $1\, \GeV$ the $\r$ 
resonance has a strong influence on the asymmetry. This can be seen from  
\fig{acpSM}, 
%
%
where we show the $\cp$ asymmetry $A_{\cp}^S$ 
[\eq{def:CPasymSD}] between $B^-\to \p^- l^+l^-$ and
$B^+\to \p^+ l^+l^-$ as a function of the dilepton invariant mass 
within the SM and in the presence of SUSY contributions with new 
$\cp$-violating phases.
It is evident that the predictions for $\cp$ asymmetries suffer from 
large theoretical uncertainties in the neighbourhood of the $\rho$ 
resonance and above the $J/\psi$, as discussed in 
\Sec{rareBdecays}.

Using Eqs.~(\ref{diff:BtoP}) and (\ref{diff:BtoV}) together with the 
definition for $\cp$-violating asymmetries, Eqs.~(\ref{def:CPasym}), 
we can summarize our main findings as follows:

\subsubsection{CP violation in $B\to P l^+l^-$}
\begin{figure}
\begin{center}
\epsfig{figure=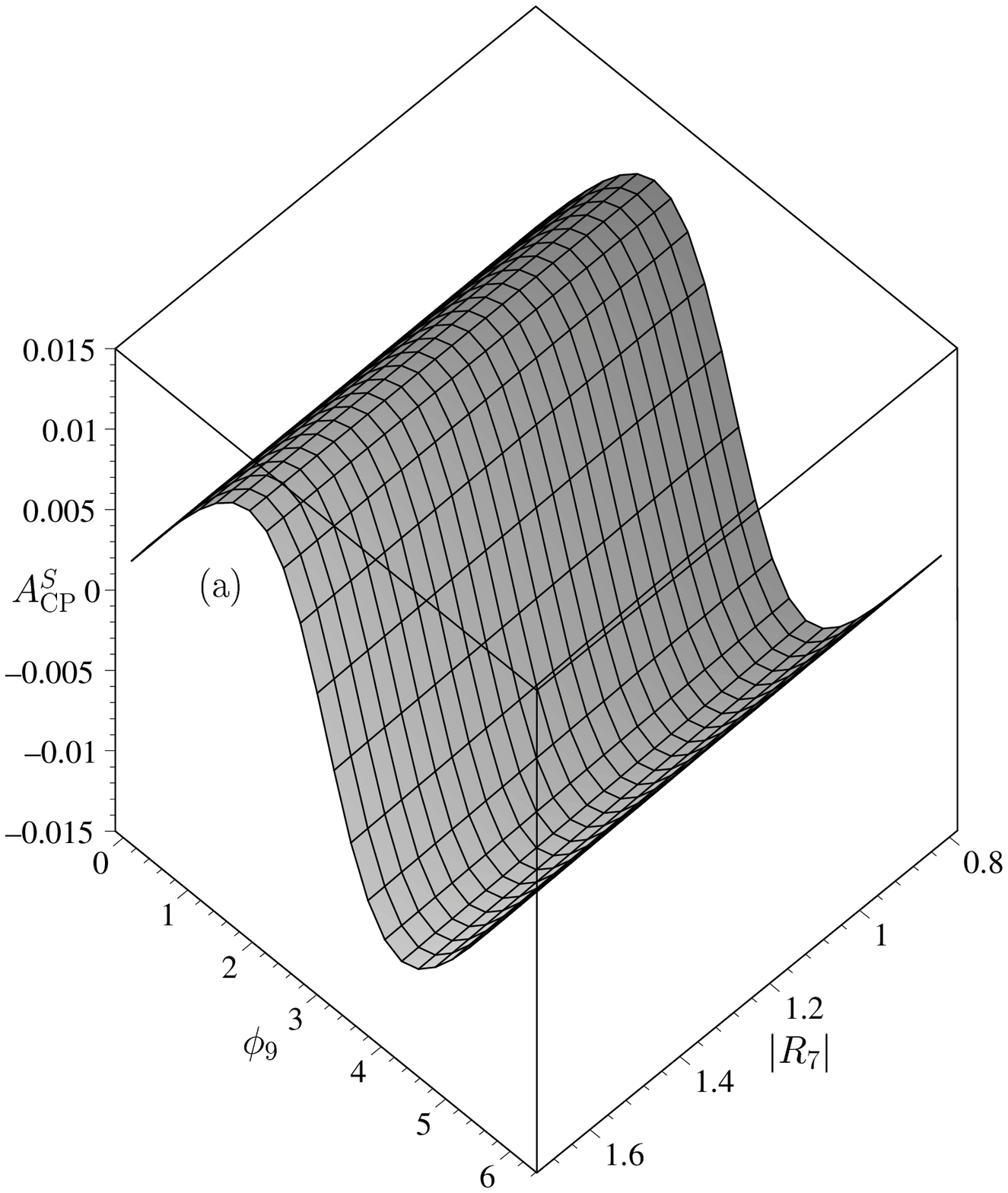,height=3.9in,angle=0}\vspace{1em}
\epsfig{figure=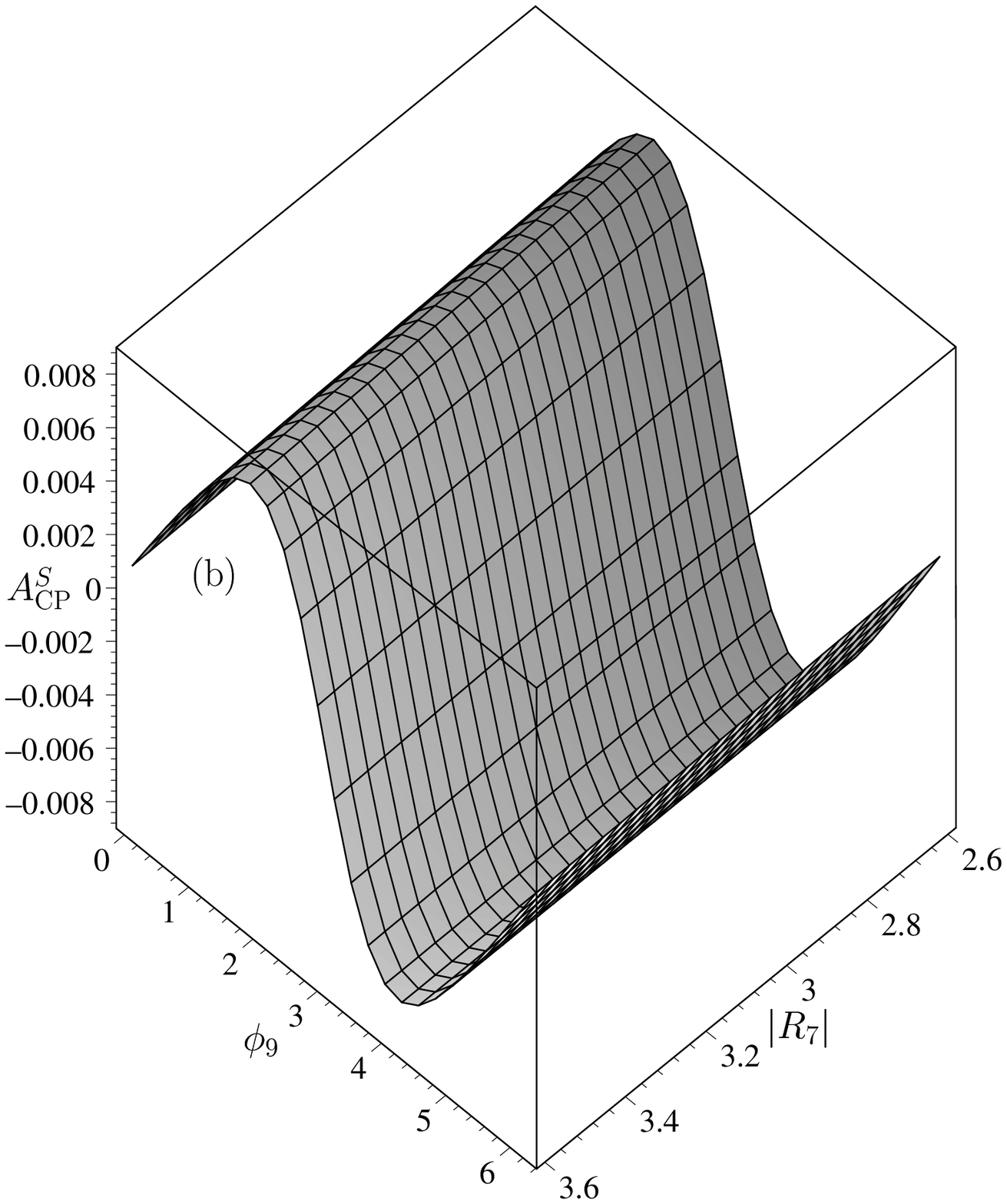,height=3.9in,angle=0}
\end{center}
\caption[]{$\cp$-violating partial width asymmetry $A_{\cp}^S$ in the decays 
$\B\to K l^+l^-$ and $B\to \bar{K} l^+l^-$ as a function of $\phi_9$ and 
$|R_7|$ for a dilepton invariant mass of $s=4\,\GeV^2$ within effective SUSY. 
(a) $\tan\b =2$ with $\phi_7=0.4$, $|R_9|=0.96$, $|R_{10}|=0.8$. (b) 
$\tan\b =30$ with $\phi_7=2.5$, $|R_9|=0.99$, $|R_{10}|=0.9$.\label{acpSeff-kdecay}}
\end{figure}

\begin{figure}
\begin{center}
\epsfig{figure=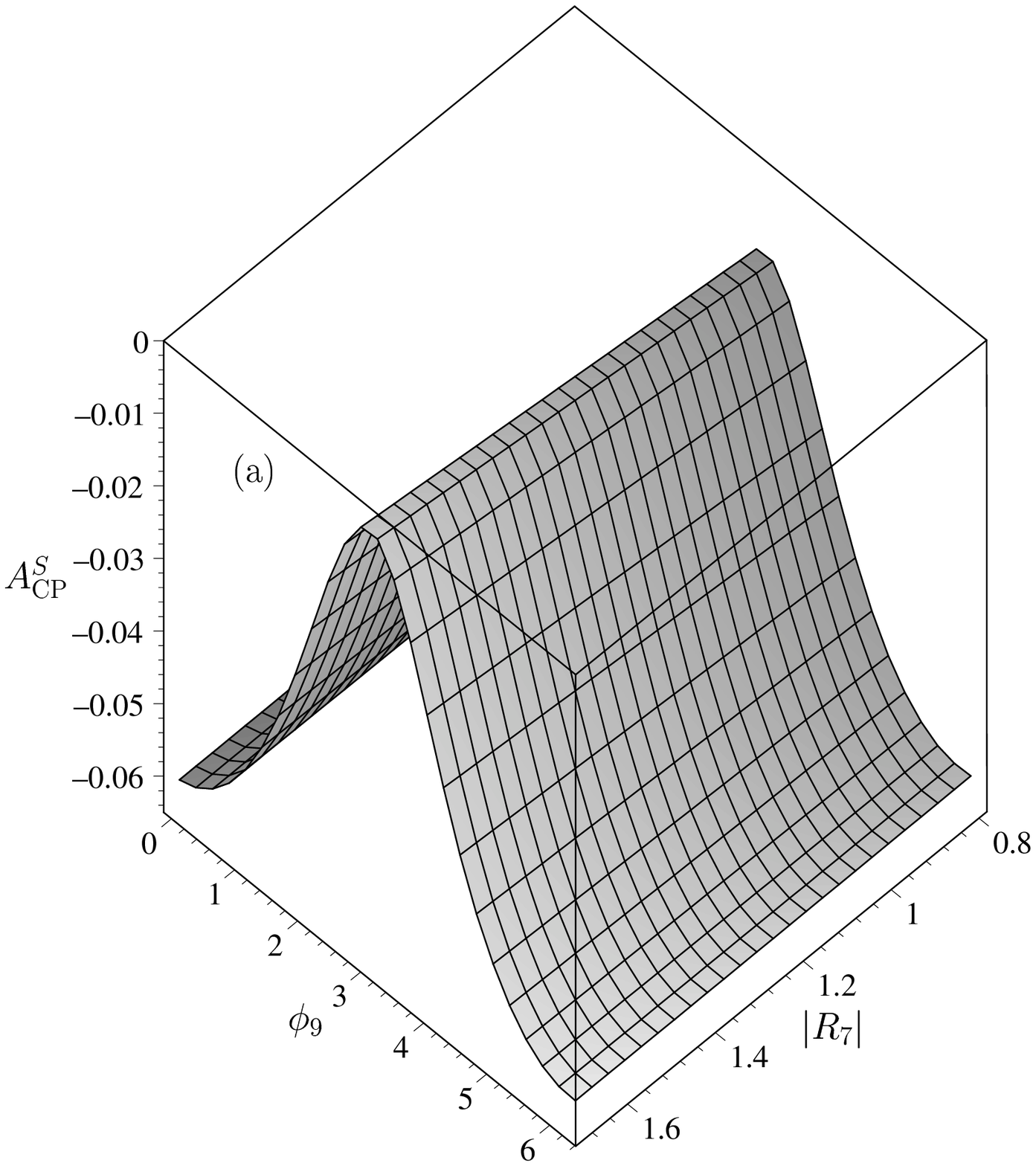,height=3.9in,angle=0}\vspace{1em}
\epsfig{figure=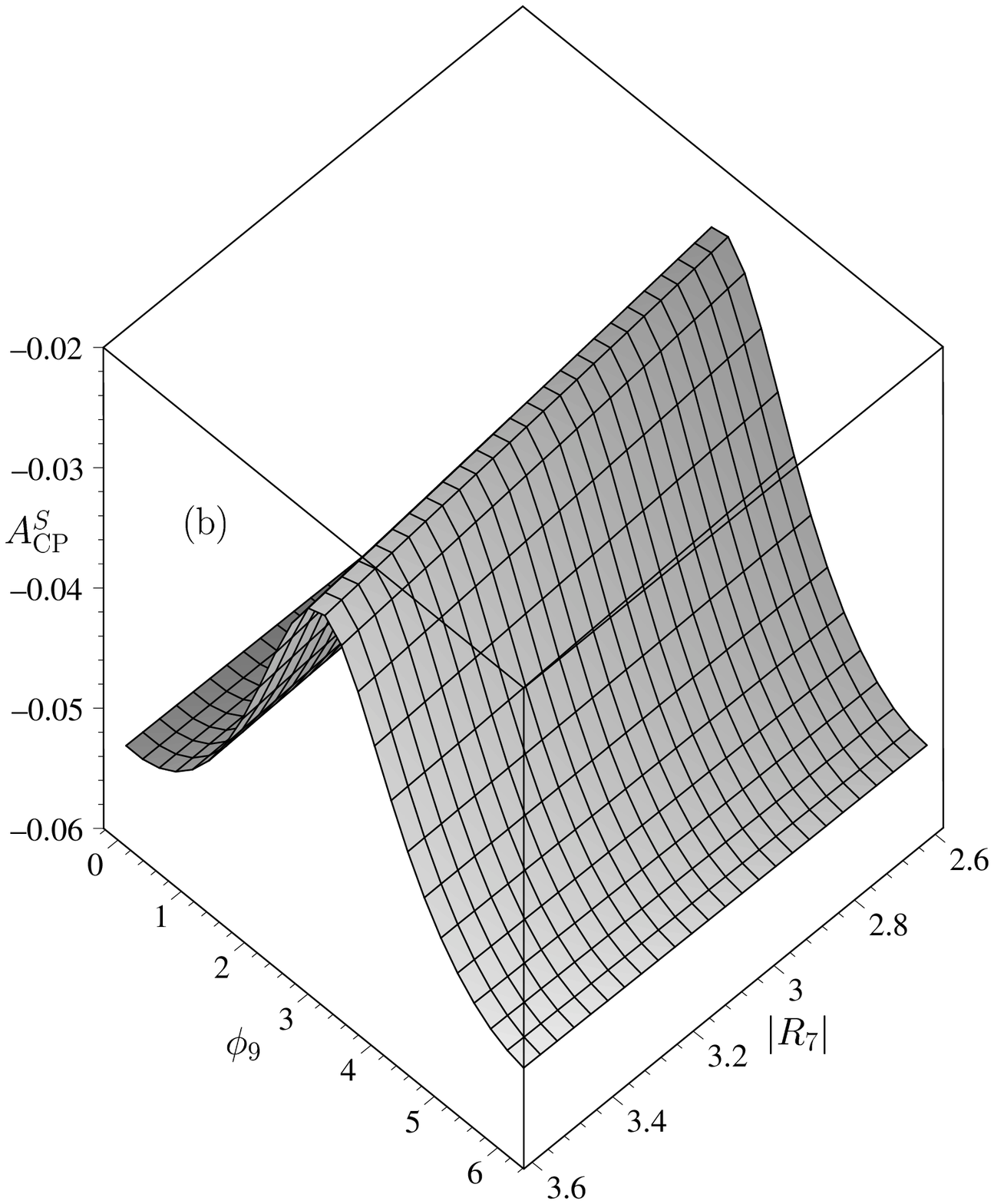,height=3.9in,angle=0}
\end{center}
\caption[]{$\cp$-violating partial width asymmetry $A_{\cp}^S$ between
$B^-\to \p^- l^+l^-$ and $B^+\to \p^+ l^+l^-$ as a function of $\phi_9$ and
$|R_7|$ for $s=4\,\GeV^2$ within effective SUSY. (a) $\tan\b =2$ with 
$\phi_7=0.4$, $|R_9|=0.96$, $|R_{10}|=0.8$. (b) $\tan\b =30$ with 
$\phi_7=2.5$, $|R_9|=0.99$, $|R_{10}|=0.9$.\label{acpSeff-pidecay}}
\end{figure}

\bit
\item The $\cp$-violating asymmetry in the $l^-$ spectra 
of $B$ and $\B$ decays, $A_{\cp}^D$, 
vanishes in the case of $B\to P$ transitions. 
Within the framework of the constrained MSSM with phases of 
${\mathcal O}(10^{-2})$ numerical values for the average asymmetry 
$\av{A_{\cp}^S}$ in the low-$s$ region are comparable to the SM predictions
with asymmetries of $0.1\%$ and $-5\%$ for $b\to s$ and $b\to d$ 
transitions respectively. 
\item Our results for $A_{\cp}^S$ between the decays 
$\B\to P l^+l^-$ and $B\to\bar{P} l^+l^-$ in the context of effective SUSY 
with a light stop 
$\tilde{t}_1$ and phases of ${\mathcal O} (1)$ are  shown in 
Figs.~\ref{acpSeff-kdecay} and \ref{acpSeff-pidecay} for low and large 
$\tan\b$ solutions which correspond to $\Re R_7 >0$ and $\Re R_7 < 0$ 
respectively. 
Observe that the $\cp$-violating asymmetry $A_{\cp}^S$ in 
$B\to P$ depends only weakly on the sign and phase of 
$\cseff$. This is due to the fact that $\cseff$, which is 
constrained by the $b\to s \g$ measurement and not 
enhanced by a factor of $1/s$ in the low-$s$ region, is nearly one order of 
magnitude smaller than the leading term in $\ceff$ [cf.~\eq{wilson:c9app}]. 
\item The $\cp$ asymmetry in the partial widths of $\B\to K l^+l^-$ and 
$B\to\bar{K} l^+l^-$ changes sign for large values of 
$\phi_9$, while $|A_{\cp}^S|\leqslant 1\%$ (see \fig{acpSeff-kdecay}).
%
%
However, non-standard contributions to $\phi_9$ are found to be small
and hence $A_{\cp}^S\sim {\mathcal O}(10^{-3})$.
On the other hand, average asymmetries of $-(5$$-$$6)\%$ are predicted 
for $\av{A_{\cp}^S}$ in the case of $B\to \p$, even for values of 
$\phi_9$ as small as $10^{-2}$ (see \fig{acpSeff-pidecay}).
%
%
Given a typical branching ratio of $10^{-8}$ and a nominal asymmetry of $6\%$, a measurement at 
$3\s$ level requires $3\times 10^{11}$ $b\bar{b}$ pairs. (This rather 
challenging task might be feasible at LHC and the Tevatron.) 
\item The small magnitude of the $\cp$ asymmetry
is also due to a suppression factor multiplying the indispensable
absorptive part in $\ceff$, which is only slightly affected by new-physics 
contributions. Indeed, it follows from 
Eqs.~(\ref{wilson:c9app}) and (\ref{def:CPasym:gen}) that  
\be
A_{\cp}\propto  r\sin\phi\sin\d \sim\frac{(3c_1+c_2)}{|c_9|}\sin\phi\sin\d
\sim 10^{-2}\sin\phi\sin\d,
\ee
where the weak and strong phases $\phi$ and $\d$ can be of order unity.
\item Numerical estimates for average $\cp$ asymmetries are 
mildly affected by the para\-me\-tri\-za\-tion of form factors 
(see also \rfs{fklms:exc,mel:np}).
\eit

\subsubsection{CP violation in $B\to V l^+l^-$}
\begin{figure}
\begin{center}
\epsfig{figure=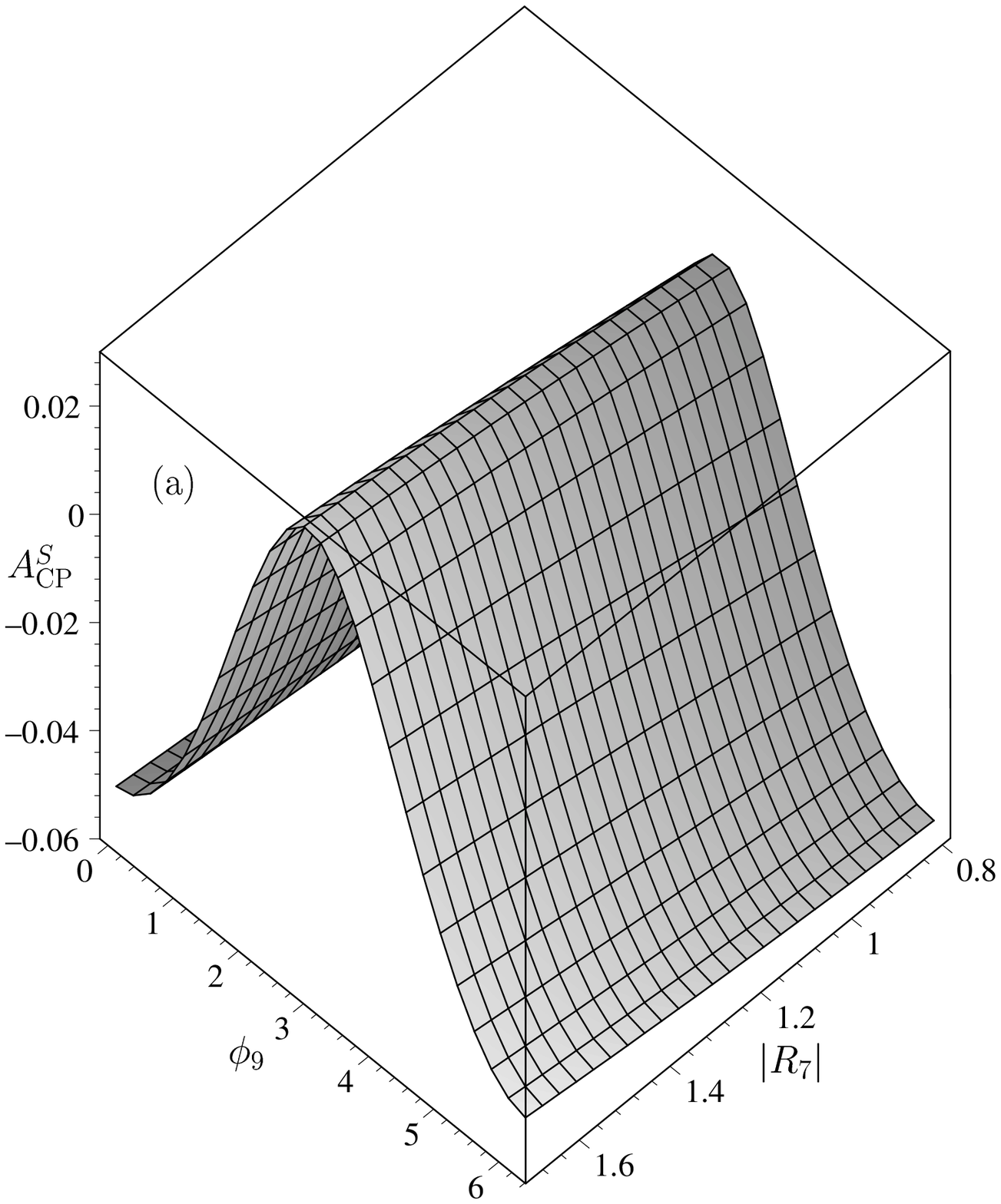,height=3.9in,angle=0}\vspace{0.5em}
\epsfig{figure=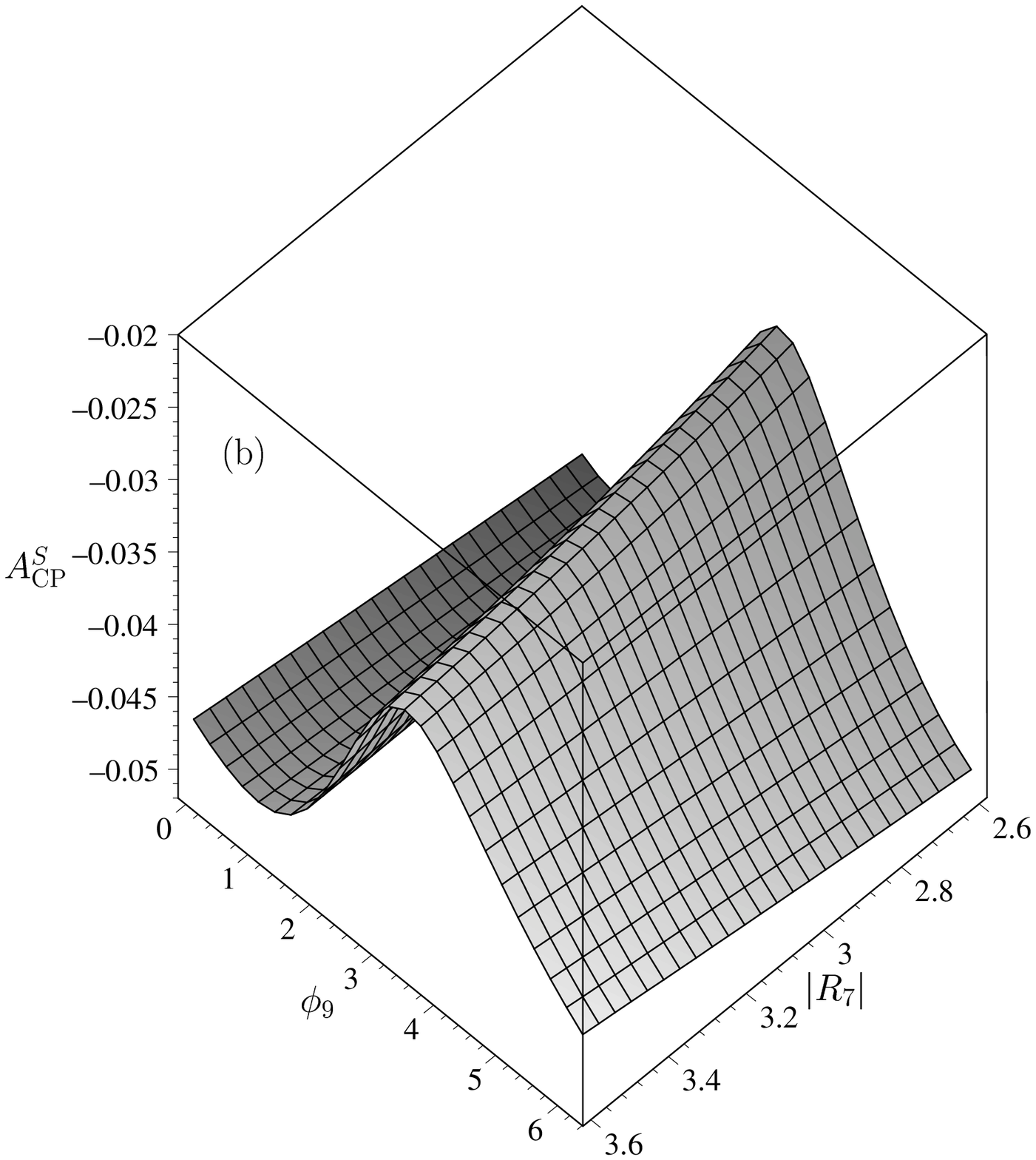,height=3.9in,angle=0}
\end{center}
\caption[]{$\cp$ asymmetry $A_{\cp}^S$ in the decays $B^-\to \r^- l^+l^-$ and 
$B^+\to \r^+ l^+l^-$ as a function of $\phi_9$ and $|R_7|$ for a dilepton 
invariant mass of $s=4\,\GeV^2$ within effective SUSY. (a) $\tan\b =2$ with 
$\phi_7=0.4$, $|R_9|=0.96$, $|R_{10}|=0.8$. (b) $\tan\b =30$ with 
$\phi_7=2.5$, $|R_9|=0.99$, $|R_{10}|=0.9$. Note that (a) and (b) 
correspond to $\Re R_7 > 0$ and $\Re R_7 < 0$ respectively.\label{acpSeff-rhodecay}}
\end{figure}

\begin{figure}
\begin{center}
\epsfig{figure=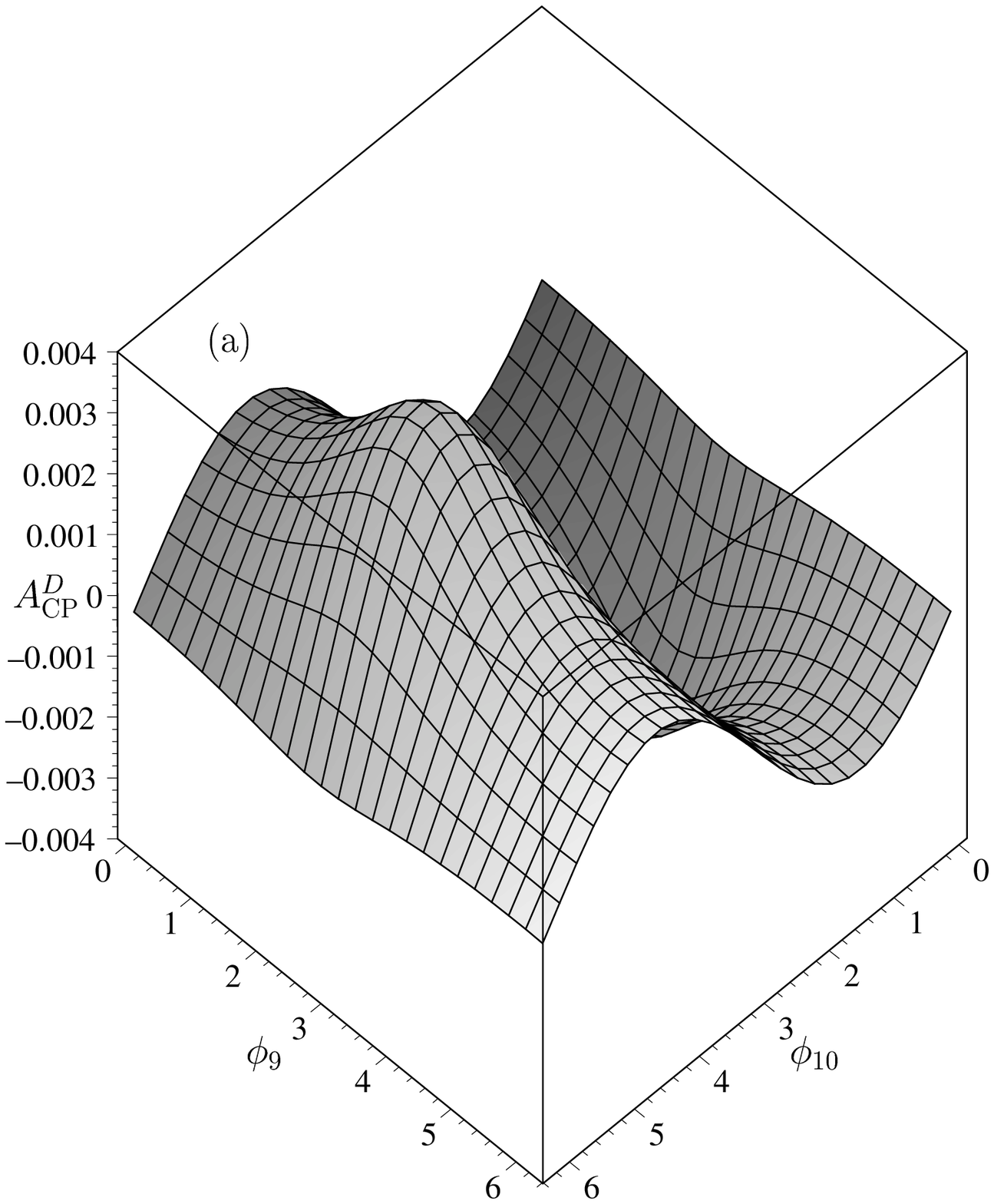,height=3.9in,angle=0}\vspace{0.5em}
\epsfig{figure=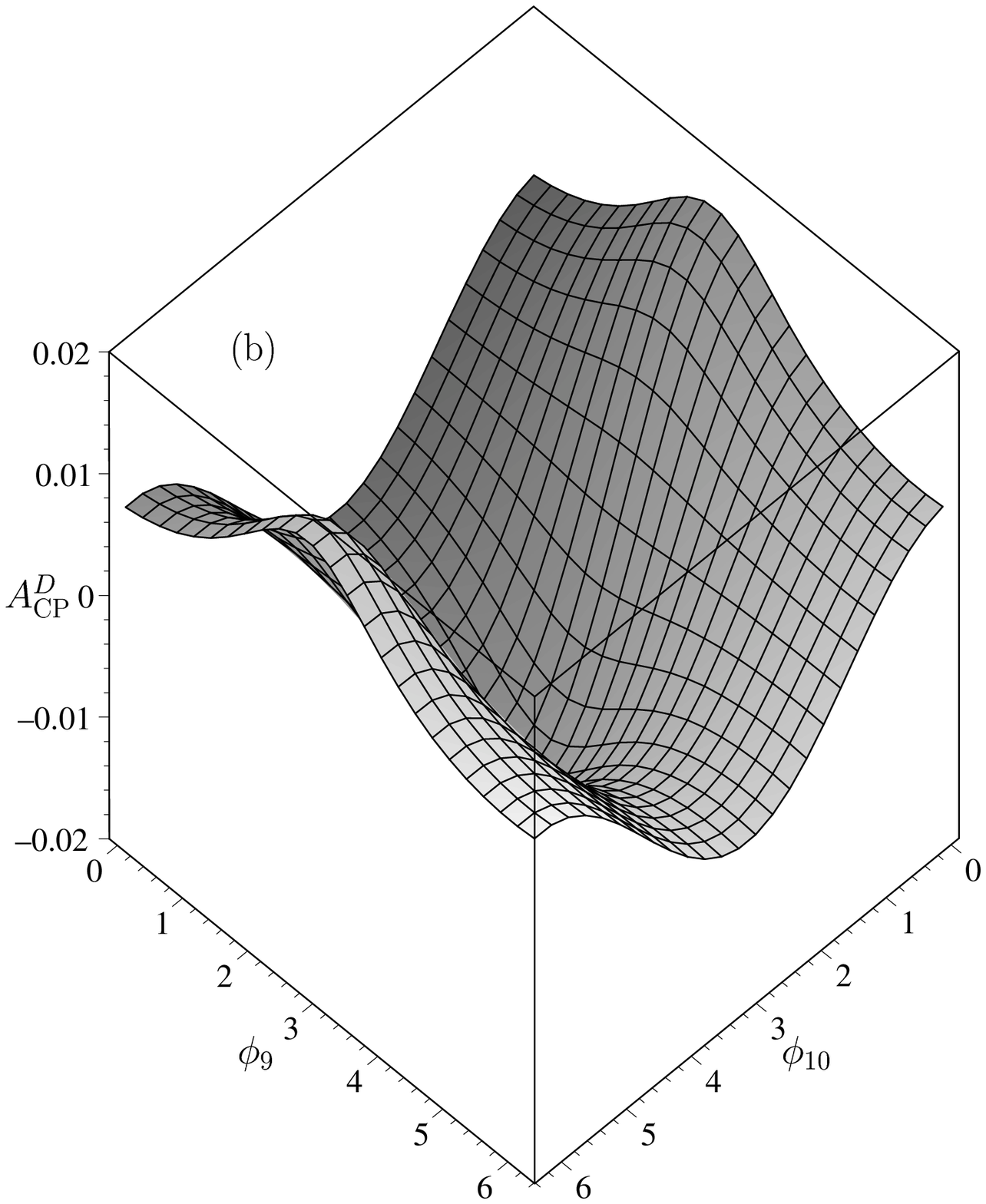,height=3.9in,angle=0}
\end{center}
\caption[]{$\cp$-violating asymmetry $A_{\cp}^D$ between (a) 
$\B\to K^* l^+l^-$ and $B\to \bar{K}^* l^+l^-$, and
(b) $B^-\to \r^- l^+l^-$ and $B^+\to \r^+ l^+l^-$ 
for large $\tan\b$ as in \fig{acpSeff-kdecay}.\label{acpD-rhodecay}}
\end{figure}

\bit
\item The contribution of the Wilson 
coefficient $\cseff$ (or equivalently $R_7$) to the decay rate in $B\to V$ 
modes is enhanced by a factor of $1/s$ in the low-$s$ region. 
As seen from \fig{acpSeff-rhodecay}, 
%
%
in the case of $B\to \r$, the $\cp$-violating asymmetry $A_{\cp}^S$ can 
change sign for $\tan\b=2$ 
(i.e.~$\Re R_7>0$), while for large $\tan\b$ it is always  
negative. For small values of $\phi_9$ an average $\cp$ asymmetry of about 
$-5\%$ is predicted for both $\tan\b=2$ and $\tan\b=30$ solutions. 
Since the distributions of $A_{\cp}^S$ for 
$B\to K^*$ are very similar to the ones 
obtained for $B\to K$, we refrain from showing the corresponding plots.
\item As we have already mentioned, the $\cp$-violating asymmetry in the 
angular distribution of $l^-$ in $B$ and $\B$ decays can, in principle, 
probe the phase of the Wilson coefficient $c_{10}$. 
This is shown in \fig{acpD-rhodecay}, 
%
%
where we have plotted the $\cp$ asymmetry as a function of 
$\phi_9$ and $\phi_{10}$ for large $\tan\b$ (i.e.~$\Re R_7 < 0$). 
Unfortunately, the mSUGRA and effective SUSY predictions for the 
average asymmetry $\av{A_{\cp}^D}$ turn out to be unobservably small. 
\eit

\section{Discussion and conclusions}\label{conclusions}
In this paper, we have studied the consequences of new $\cp$-violating 
phases for exclusive $B$ decays within the framework of 
supersymmetric extensions of the SM, ignoring intergenerational mixing in 
the squark sector. We have examined $\cp$-violating asymmetries in 
the partial widths as well as 
angular distribution of $l^-$ between the exclusive channels 
$b\to s(d)l^+l^-$ and $\bar{b}\to \bar{s}(\bar{d}) l^+l^-$ in the 
invariant mass region $1.2\, \GeV < M_{l^+l^-}< 2.9\,\GeV$. 
The essential conclusion of our analysis is that it is not 
sufficient to have additional $\cp$ phases of ${\mathcal O}(1)$ in order 
to obtain large $\cp$-violating effects. 

Within the constrained MSSM and effective SUSY with a complex 
CKM matrix and additional $\cp$ phases, we obtain values for 
the average asymmetry $\av{A_{\cp}^S}$ of about $-6\%$ ($-5\%$) 
in the decays $\B\to \p (\r) l^+l^-$ and $B\to \bar{\p} (\bar{\r}) l^+l^-$, 
taking into account experimental constraints on 
EDM's of electron and neutron, as well as rare $B$ decays such as 
$b\to s\g$. As for the asymmetry in the angular distribution, 
$\av{A_{\cp}^D}$, it probes the phase of the Wilson coefficient 
$c_{10}$, but will be unobservable at future colliders. 
Numerical estimates of the $\cp$ asymmetries in the decays 
$\B\to K^{(*)}l^+l^-$ and 
$B\to \bar{K}^{(*)}l^+l^-$ turn out to be small 
(less than $1\%$) and are comparable to the SM result. 

Our analysis shows that the smallness in $\cp$ asymmetries 
is mainly due to the coefficient $(3c_1+c_2)/|c_9|$
which multiplies the requisite absorptive part in $\ceff$ 
[\eq{wilson:c9app}], 
and which is only slightly affected by the
new-physics contributions discussed in \Sec{rareBdecays}. 
Therefore, any sizable 
$\cp$-violating effect in the low-$s$ region requires large non-standard 
contributions to the short-distance coefficient $\ceff$ and/or $c_{10}$,
as well as additional $\cp$ phases of ${\mathcal O}(1)$.~By the same token, 
any large $\cp$-violating effect would provide a clue to physics beyond the 
SM. A detailed discussion of this point will be given elsewhere. 

One could argue, however, that the inclusion of flavour off-diagonal 
contributions (i.e.~gluino and neutralino diagrams) to the 
Wilson coefficients might lead to higher $\cp$ asymmetries. 
In fact, it has been pointed out in \rfs{offdiag:cmssm,flavour-structure} 
that even in the presence of 
large supersymmetric $\cp$ phases, a non-trivial 
flavour structure in the soft-breaking terms is 
necessary in order to obtain sizable contributions to $\cp$ violation 
in the $K$ system and to $\cp$ asymmetries in two-body 
neutral $B$ decays (see also \rf{phase:non-uni}). 
Using the mass insertion approximation, such effects have recently been 
studied in \rf{enrico:recent} which predicts a partial width asymmetry for 
$b\to sl^+l^-$ of a few per cent in the low-$s$ domain.

Finally, we would like to recall that, nevertheless, large $\cp$ 
asymmetries may occur in rare $B$ decays like the 
observed $b\to s\g$ modes, where $A_{\cp}$ can be substantial 
(up to $\pm 45\%$) in some part of the 
parameter space \cite{off-diagonal,kagan:neubert:cp}.

\acknowledgments
F.K. would like to thank Gudrun Hiller for useful discussions.
This research has been supported by the TMR Network of the EC under 
contract ERBFMRX-CT96-0090.

\appendix
\section{Numerical inputs}\label{app:numerics}
Unless otherwise specified, we use the experimental values as compiled by 
the Particle Data Group \cite{pdg} and 
the parameters displayed in \eq{eqs:input}.
\bea\label{eqs:input}
\begin{array}{l}
m_t=175\,\GeV,\ m_b=4.8\,\GeV,\ m_c=1.4\,\GeV,\ m_s=170\, \MeV,
\ m_d=10\,\MeV,\\
m_u=5\,\MeV,\ \a=1/129,\ \Lambda_{\text{QCD}}=225\,\MeV.
\end{array}
\eea

\section{Wilson coefficients and SUSY}\label{app:wilson}
For the sake of convenience, we provide in this appendix formulae for the 
Wilson coefficients $\cseff$, $\ceff$, and $c_{10}$ in the presence of SUSY, 
using the results derived in \rfs{bertolini:etal,cho:etal,bmm,ciuchini:etal}. 
Since we study the case of massless leptons, 
we retain only those contributions that do 
not vanish in the limit $m_l\to 0$. As for $\t$ leptons in the final state, 
there are further charged and neutral Higgs-boson contributions [see also 
Eqs.~(\ref{feyn:X})]. 

Introducing the shorthand notation
\be\label{notation}
\eta_s=\frac{\a_s(M_W)}{\a_s(m_b)}, \quad r_W= \frac{m_t^2}{M_W^2},
\quad r_{\higgs}=  \frac{m_t^2}{m_{\higgs}^2},
\quad \R{A}{B}= \frac{m_A^2}{m_B^2},
\ee
the Wilson coefficient $\cseff$ evaluated at $\m_R=m_b$
has the form (in leading-log approximation)   
\be\label{wilson:c7eff}
\cseff=\eta_{s}^{16/23}c_7(M_W)+\frac{8}{3}(\eta_{s}^{14/23}-\eta_{s}^{16/23})c_8(M_W)+\sum_{i=1}^8 h_i\eta_{s}^{a_i},
\ee 
with the coefficients $a_i$, $h_i$ tabulated in \rf{bmm}. 
Recalling Eqs.~(\ref{feyn:X}) and (\ref{wilson:komplett}), and using the 
one-loop functions $f_i$ listed in Appendix \ref{app:auxfuncs}, the various 
contributions to $c_{7,8}(M_W)$ can be written as follows:
\bit
\item Standard model: 
\eit
\be\label{c7:sm}
c_7^{\sm}(M_W)= \frac{1}{4}r_W f_1(r_W).
\ee
%
%
\bit
\item Charged Higgs boson:
\eit
\be
c_7^{\higgs}(M_W)=\frac{1}{12}[r_{\higgs}
f_1(r_{\higgs})\cot^2\b + 2 f_2(r_{\higgs})].
\ee
\bit
\item Chargino:\footnote{Notice that the one-loop function 
appearing in the last term of \eq{c7mw:chargino} is actually $f_2 + 5/2$. 
However, using the explicit form for the 
squark mixing matrices [Eqs.~(\ref{feyn:X})], the constant term vanishes
-- reflecting the unitarity of the mixing matrices.}
\eit
\bea\label{c7mw:chargino}
\lefteqn{c_7^{\chargino}(M_W)
=-\frac{1}{6g^2V_{tb}^{}V_{tf}^*}\sum_{a=1}^6\sum_{j=1}^2
\frac{M_W^2}{m^2_{\charginoi{j}}}}\nnu\\ 
&\times&\Bigg[(X_j^{U_L\dagger})_{na}(X_j^{U_L})_{a3}^{}f_1(\R{\squarkui{a}}{\charginoi{j}}) 
-2 (X_j^{U_L\dagger})_{na}(X_j^{U_R})_{a3}^{}\frac{m_{\charginoi{j}}}{m_b}
f_2(\R{\squarkui{a}}{\charginoi{j}})\Bigg],
\eea
where we have defined 
\be
n=\left\{\begin{array}{l}1\quad \text{for}\quad f=d,\\ 
2\quad \text{for}\quad f=s.\end{array}\right.
\ee

For completeness, we also give the expressions for the 
neutralino and gluino contributions which vanish in 
the limit of flavour-diagonal squark mass matrices. 
\bit
\item Neutralino:
\eit
\bea\label{c7:neutralino}
\lefteqn{c_7^{\neutralino}(M_W)}\nnu\\
&=&-\frac{1}{6g^2V_{tb}^{}V_{tf}^*}\sum_{a=1}^6\sum_{k=1}^4 
\frac{M_W^2}{m^2_{\neutralinoi{k}}}
\Bigg[(Z_k^{D_L\dagger})_{na}(Z_k^{D_L})_{a3}^{}f_3(\R{\squarkdi{a}}{\neutralinoi{k}}) +2(Z_k^{D_L\dagger})_{na}
(Z_k^{D_R})_{a3}^{}\frac{m_{\neutralinoi{k}}}{m_b}
f_4(\R{\squarkdi{a}}{\neutralinoi{k}})\Bigg].
\eea
\bit
\item Gluino:
\eit
\be
c_7^{\gluino}(M_W)=
-\frac{4g_s^2}{9g^2V_{tb}^{}V_{tf}^*}\sum_{a=1}^6
\frac{M_W^2}{m^2_{\gluino}}
\Bigg[(G^{D_L\dagger})_{na}(G^{D_L})_{a3}^{}
f_3(\R{\squarkdi{a}}{\gluino}) 
-2 (G^{D_L\dagger})_{na}(G^{D_R})_{a3}^{}\frac{m_{\gluino}}{m_b}
f_4(\R{\squarkdi{a}}{\gluino})\Bigg].
\ee

The corresponding expressions $c_8^{\sm}(M_W), \dots, c_8^{\neutralino}(M_W)$ 
are obtained changing $f_i\to g_i$ in \eqs{c7:sm}{c7:neutralino}, 
with $g_i$ collected in Appendix \ref{app:auxfuncs}, while the gluino contribution reads
\be
c_8^{\gluino}(M_W)=
-\frac{4g_s^2}{9g^2V_{tb}^{}V_{tf}^*}\sum_{a=1}^6
\frac{M_W^2}{m^2_{\gluino}}
\Bigg[(G^{D_L\dagger})_{na}(G^{D_L})_{a3}^{}g_5(\R{\squarkdi{a}}{\gluino}) 
-2 (G^{D_L\dagger})_{na}(G^{D_R})_{a3}^{}\frac{m_{\gluino}}{m_b}
g_6(\R{\squarkdi{a}}{\gluino})\Bigg].
\ee
%
 
The Wilson coefficient $\ceff$ at $\m_R=m_b$ in next-to-leading 
approximation is given by
\bea\label{wilson:c9}
\ceff&=&c_9 \left[ 1 + \frac{\a_s(m_b)}{\p}\omega(s/m_b^2)\right] + g(m_c,s) \left(3 c_1 + c_ 2 +3c_3 + c_4 + 3c_5 + c_6\right)\nnu\\ 
& +& \l_u\left[ g(m_c,s)- g(m_u,s)\right]\left(3 c_1 + c_ 2 \right)  
-\frac{1}{2} g(m_f,s) \left(c_3 + 3 c_4\right)
\nnu\\
& -&\frac{1}{2} g(m_b,s)\left(4c_3 + 4c_4 +3c_5+c_6\right)+ 
\frac{2}{9}\left(3c_3 +c_4+3c_5+c_6\right),
\eea
where $\l_u$ and $g(m_i, s)$ are defined in 
Eqs.~(\ref{lambdau}) and (\ref{loopfunc}) respectively, with
$s\equiv q^2$. As far as the Wilson coefficients 
$c_1$--$c_6$ are concerned, we have numerically
\be
c_1=-0.249,\ c_2=1.108,\ c_3=0.011,\ c_4=-0.026,\ c_5=0.007,\ c_6=-0.031,
\ee
using the values given in Appendix \ref{app:numerics}.
Further, 
\be\label{c9}
c_9= c_9(M_W)-\frac{4}{9}+ P_0 + P_E \sum_i E^i,  
\ee
with $i={\sm}, \higgs, \chargino, \neutralino, \gluino$, and 
\be\label{c9mw}
c_9(M_W)= \sum_i\Bigg(\frac{Y^i}{\sin^2\theta_W}-4Z^i\Bigg)+\frac{4}{9},
\ee
where the analytic expressions for $P_0$, $P_E$, and $E^{\sm}$ 
are given in \rf{bmm}. Since $P_E\ll P_0$, we shall keep only the SM 
contribution 
in the last term of \eq{c9}. Moreover, as discussed in \rf{lw}, 
the order $\a_s$ correction in \eq{wilson:c9} due 
to one-gluon exchange may be regarded as a contribution to the form factors, 
and hence we set $\omega =0$ in \eq{wilson:c9}.

Turning to the Wilson coefficient $c_{10}$, it has the simple form    
\be
c_{10}(M_W) = -\sum_i\frac{Y^i}{\sin^2\theta_W}.
\ee 
Note that the corresponding operator does 
not renormalize and thus  $c_{10}(M_W)=c_{10}(\m_R)$. 

The expressions for the various contributions to $Y$ and $Z$ read as 
follows:\footnote{Regarding the expressions for the chargino and 
neutralino box-diagram contributions, and the sign discrepancy between 
\rf{bertolini:etal} and \rf{cho:etal}, we confirm the results of the latter.}
\bit
\item Standard model:
\eit
\be
Y^{\sm}= \frac{1}{8}f_9(r_W), \quad Z^{\sm}= \frac{1}{72}f_{10}(r_W).
\ee
\bit
\item Charged Higgs:
\eit
\begin{mathletters}
\be
Y^{\higgs}\equiv Y^{\higgs}_Z, \quad 
Z^{\higgs}\equiv Z^{\higgs}_{\g}+Z^{\higgs}_Z,
\ee
\be
Z^{\higgs}_{\g}=-\frac{1}{72}f_6(r_{\higgs})\cot^2\b, 
\quad 
Y^{\higgs}_Z= Z^{\higgs}_Z=-\frac{1}{8}r_W f_5(r_{\higgs})\cot^2\b.
\ee
\end{mathletters}
\bit
\item Chargino:
\eit
\begin{mathletters}
\be
Y^{\chargino}\equiv Y^{\chargino}_Z+ Y^{\chargino}_{\text{box}}, \quad 
Z^{\chargino}\equiv Z^{\chargino}_{\g}+Z^{\chargino}_Z,
\ee
\be
Z^{\chargino}_{\g}=\frac{1}{36g^2V_{tb}^{}V_{tf}^*}
\sum_{a=1}^6 \sum_{j=1}^2\frac{M_W^2}{m^2_{\squarkui{a}}}[ 
(X_j^{U_L\dagger})_{na}(X_j^{U_L})_{a3}f_7(\R{\charginoi{j}}{\squarkui{a}})], 
\ee
\bea
\lefteqn{Y^{\chargino}_Z= Z^{\chargino}_Z=
\frac{1}{2g^2V_{tb}^{}V_{tf}^*}\sum_{a,b= 1}^6 \sum_{i,j= 1}^2
\Bigg\{(X_i^{U_L\dagger})_{na}(X_j^{U_L})_{b3}
\Bigg[c_2(m^2_{\charginoi{i}}, 
m^2_{\squarkui{a}}, m^2_{\squarkui{b}})(\G^{U_L}\G^{U_L\dagger})_{ab} 
\d_{ij}}\nnu\\
&-&\mbox{}c_2(m^2_{\squarkui{a}},m^2_{\charginoi{i}},m^2_{\charginoi{j}})
\d_{ab}V^*_{i1}V_{j1}^{}+\frac{1}{2}m_{\charginoi{i}}m_{\charginoi{j}}
c_0(m^2_{\squarkui{a}}, m^2_{\charginoi{i}}, m^2_{\charginoi{j}})\d_{ab}
U_{i1}^{}U^*_{j1}\Bigg]\Bigg\},
\eea
\be
Y^{\chargino}_{\text{box}}=\frac{M_W^2}{g^2V_{tb}^{}V_{tf}^*}
\sum_{a=1}^6 \sum_{i,j=1}^2[(X_i^{U_L\dagger})_{na}(X_j^{U_L})_{a3}
d_2(m^2_{\charginoi{i}}, m^2_{\charginoi{j}}, m^2_{\squarkui{a}}, 
m^2_{\sneutrinoi{1,2}})V_{i1}^*V_{j1}],
\ee
with $m_{\sneutrino_1}(m_{\sneutrino_2})$ in the case of 
$e^+e^-(\m^+\m^-)$ in the final state. 
\end{mathletters}
\bit
\item Neutralino:
\eit
\begin{mathletters}
\be
Y^{\neutralino}\equiv Y^{\neutralino}_Z+ Y^{\neutralino}_{\text{box}}, \quad 
Z^{\neutralino}\equiv Z^{\neutralino}_{\g}+Z^{\neutralino}_Z+ 
Z^{\neutralino}_{\text{box}}, 
\ee
\be
Z^{\neutralino}_{\g}=-\frac{1}{216g^2V_{tb}^{}V_{tf}^*}
\sum_{a= 1}^6\sum_{k= 1}^4\frac{M_W^2}{m^2_{\squarkdi{a}}}[
(Z_k^{D_L\dagger})_{na}(Z_k^{D_L})_{a3}
f_8(\R{\neutralinoi{k}}{\squarkdi{a}})],
\ee
\bea
\lefteqn{Y^{\neutralino}_Z= Z^{\neutralino}_Z=
\frac{1}{2g^2V_{tb}^{}V_{tf}^*}\sum_{a,b= 1}^6 \sum_{k,l= 1}^4
\Bigg\{(Z_k^{D_L\dagger})_{na}(Z_l^{D_L})_{b3}}\nnu\\
&\times &\Bigg[c_2(m^2_{\neutralinoi{k}}, 
m^2_{\squarkdi{a}}, m^2_{\squarkdi{b}})(\G^{D_R}\G^{D_R\dagger})_{ab} 
\d_{kl}
- c_2(m^2_{\squarkdi{a}}, m^2_{\neutralinoi{k}}, m^2_{\neutralinoi{l}})
\d_{ab}(N^*_{k3}N_{l3}^{}-N^*_{k4}N_{l4}^{})\nnu\\
&&\hspace{0.3em}\mbox{}-\frac{1}{2}m_{\neutralinoi{k}}m_{\neutralinoi{l}}
c_0(m^2_{\squarkdi{a}}, m^2_{\neutralinoi{k}}, m^2_{\neutralinoi{l}})\d_{ab}
(N_{k3}^{}N^*_{l3}-N_{k4}^{}N^*_{l4})\Bigg]\Bigg\},
\eea
\bea
\lefteqn{Y^{\neutralino}_{\text{box}}= 
2\sin^2\theta_W Z^{\neutralino}_{\text{box}}
+ \frac{M_W^2}{2g^2V_{tb}^{}V_{tf}^*}\sum_{a=1}^6\sum_{k,l=1}^4\Bigg\{
(Z_k^{D_L\dagger})_{na}(Z_l^{D_L})_{a3}}\nnu\\
&\times &\Bigg[
d_2(m^2_{\neutralinoi{k}}, m^2_{\neutralinoi{l}}, m^2_{\squarkdi{a}}, 
m^2_{\sleptoni{1,2}})(N_{k2}^*+ \tanw N_{k1}^*)(N_{l2}+ \tanw N_{l1})\nnu\\
&&\hspace{0.3em}\mbox{}+\frac{1}{2}m_{\neutralinoi{k}}m_{\neutralinoi{l}}
d_0(m^2_{\neutralinoi{k}}, m^2_{\neutralinoi{l}}, m^2_{\squarkdi{a}}, 
m^2_{\sleptoni{1,2}})(N_{k2}+\tanw N_{k1})(N_{l2}^*+\tanw N_{l1}^*)
\Bigg]\Bigg\},
\eea
\bea
\lefteqn{Z^{\neutralino}_{\text{box}}=\frac{M_W^2}{g^2V_{tb}^{}V_{tf}^*}
\sum_{a=1}^6\sum_{k,l=1}^4\Bigg\{(Z_k^{D_L\dagger})_{na}(Z_l^{D_L})_{a3}
\sec^2\theta_W} \nnu\\
&\times&\Bigg[d_2(m^2_{\neutralinoi{k}}, m^2_{\neutralinoi{l}}, 
m^2_{\squarkdi{a}}, m^2_{\sleptoni{4,5}})N_{k1}^*N_{l1}
+\frac{1}{2}m_{\neutralinoi{k}}m_{\neutralinoi{l}}
d_0(m^2_{\neutralinoi{k}}, m^2_{\neutralinoi{l}}, m^2_{\squarkdi{a}}, 
m^2_{\sleptoni{4,5}})N_{k1}N_{l1}^*\Bigg]\Bigg\},
\eea
with $m_{\sleptoni{1,4}}(m_{\sleptoni{2,5}})$ for $e^+e^-(\m^+\m^-)$ 
in the final state.
\end{mathletters}
\bit
\item Gluino:
\eit
\begin{mathletters}
\be
Y^{\gluino}\equiv Y^{\gluino}_Z, \quad 
Z^{\gluino}\equiv Z^{\gluino}_{\g}+Z^{\gluino}_Z,
\ee
\be
Z^{\gluino}_{\g}=
-\frac{g_s^2}{81g^2V_{tb}^{}V_{tf}^*}\sum_{a= 1}^6
\frac{M_W^2}{m^2_{\squarkdi{a}}}[(G^{D_L\dagger})_{na}(G^{D_L})_{a3}
f_8(\R{\gluino}{\squarkdi{a}})],
\ee
\be
Y^{\gluino}_Z=Z^{\gluino}_Z= \frac{4g_s^2}{3g^2V_{tb}^{}V_{tf}^*}
\sum_{a,b= 1}^6[(G^{D_L\dagger})_{na}(G^{D_L})_{b3}
c_2(m^2_{\gluino}, m^2_{\squarkdi{a}}, m^2_{\squarkdi{b}}) 
(G^{D_R}G^{D_R\dagger})_{ab}].
\ee
\end{mathletters}%
The functions $f_i$, $c_i$, and $d_i$ are given in \eqs{fis}{dis} below.

\section{Auxiliary functions}\label{app:auxfuncs}
Here we list the functions $f_i$, $g_i$, $c_i$, and $d_i$ introduced in 
the previous section.
\begin{mathletters}\label{fis}
\be
f_1(x)=\frac{-7+5x+8x^2}{6(1-x)^3} - \frac{x(2-3x)}{(1-x)^4}\ln x,
\ee
\be
f_2(x)=\frac{x(3-5x)}{2(1-x)^2} + \frac{x(2-3x)}{(1-x)^3}\ln x,
\ee
\be
f_3(x)=\frac{2+5x-x^2}{6(1-x)^3} + \frac{x}{(1-x)^4}\ln x,
\ee
\be
f_4(x)=\frac{1+x}{2 (1-x)^2} + \frac{x}{(1-x)^3}\ln x,
\ee
\be
f_5(x)=\frac{x}{1-x} + \frac{x}{(1-x)^2}\ln x,
\ee
\be
f_6(x)=\frac{x(38-79x+47x^2)}{6(1-x)^3} + \frac{x(4-6x+3x^3)}{(1-x)^4}\ln x,
\ee
\be
f_7(x)=\frac{52-101x+43x^2}{6(1-x)^3} + \frac{6-9x+2x^3}{(1-x)^4}\ln x, 
\ee
\be
f_8(x)=\frac{2-7x+11x^2}{(1-x)^3} + \frac{6x^3}{(1-x)^4}\ln x,
\ee
\be
f_9(x)=\frac{x(4-x)}{1-x}+\frac{3x^2}{(1-x)^2}\ln x,
\ee
\be
f_{10}(x)=\frac{x(108-259x+163x^2-18x^3)}{2(1-x)^3}
-\frac{8-50x+63x^2+6x^3-24x^4}{(1-x)^4}\ln x,
\ee
\end{mathletters}

\begin{mathletters}\label{gis}
\be
g_1(x)=-3f_3(x),\quad g_2(x)=3[f_4(x)-1/2], 
\quad g_3(x)=-3f_3(x), \quad g_4(x)=-3f_4(x),
\ee
\be
g_5(x)=\frac{1}{8}\Bigg[
\frac{11-40x-19x^2}{2(1-x)^3}+\frac{3x(1-9x)}{(1-x)^4}\ln x\Bigg],
\ee
\be
g_6(x)=\frac{3}{8}
\Bigg[\frac{5-13x}{(1-x)^2}+ \frac{x(1-9x)}{(1-x)^3}\ln x\Bigg],
\ee
\end{mathletters}

\begin{mathletters}\label{cis}
\be
c_0(m_1^2, m_2^2, m_3^2) = -\Bigg[\frac{m_1^2 \ln(m_1^2/\m_R^2)}{(m_1^2-
m_2^2)(m_1^2-m_3^2)}+ (m_1 \leftrightarrow m_2)+ (m_1 \leftrightarrow m_3)
\Bigg], 
\ee
\be\label{cis:c2}
c_2(m_1^2, m_2^2, m_3^2) = \frac{3}{8}-\frac{1}{4}\Bigg[\frac{m_1^4
\ln(m_1^2/\m_R^2)}{(m_1^2-m_2^2)(m_1^2-m_3^2)}+ (m_1 \leftrightarrow m_2)
+ (m_1 \leftrightarrow m_3)\Bigg],
\ee
\end{mathletters}

\begin{mathletters}\label{dis}
\bea
\lefteqn{d_0(m_1^2, m_2^2, m_3^2, m_4^2)}\nnu\\
&=&-\Bigg[\frac{m_1^2\ln(m_1^2/\m_R^2)}{(m_1^2-m_2^2)(m_1^2-m_3^2)(m_1^2-m_4^2)} + (m_1 \leftrightarrow m_2)+ (m_1 \leftrightarrow m_3) 
+ (m_1 \leftrightarrow m_4) \Bigg],
\eea
\bea
\lefteqn{d_2(m_1^2, m_2^2, m_3^2, m_4^2)}\nnu\\
&=&- \frac{1}{4}\Bigg[\frac{m_1^4\ln(m_1^2/\m_R^2)}{(m_1^2-m_2^2)(m_1^2-
m_3^2)(m_1^2-m_4^2)} 
+ (m_1 \leftrightarrow m_2)
+ (m_1 \leftrightarrow m_3) 
+ (m_1 \leftrightarrow m_4) \Bigg].
\eea
\end{mathletters}

\section{Form factors}\label{formfactors}
In Tables \ref{table:ff:mel} and \ref{table:ff:col} we summarize the two different sets 
of form factors discussed in \Sec{decays:exc}, which are related via
\begin{mathletters}\label{formfactors:relation}
\be
F_1(q^2)=f_+(q^2),
\ee
\be
F_0(q^2)=f_+(q^2) +\frac{q^2}{M_B^2-M_P^2}f_-(q^2),
\ee
\be
F_T(q^2)= - (M_B + M_P)s(q^2),
\ee
\be
V(q^2) = (M_B+M_V) g(q^2),
\ee
\be
A_1(q^2)= \frac{f(q^2)}{M_B+M_V},
\ee
\be
A_2(q^2)= -(M_B+M_V) a_+(q^2),
\ee
\be
A_0(q^2)=  \frac{q^2 a_-(q^2) + f(q^2) + (M_B^2-M_V^2) a_+(q^2)}{2M_V},
\ee
\be
T_1(q^2)= -\frac{1}{2} g_+(q^2),
\ee
\be
T_2(q^2)= -\frac{1}{2} \Bigg[g_+(q^2) + \frac{q^2 g_-(q^2)}{M_B^2- M_V^2}
\Bigg],
\ee
\be
T_3(q^2)= \frac{1}{2}\Bigg[g_-(q^2) - \frac{(M_B^2- M_V^2)h(q^2)}{2}\Bigg] .
\ee
\end{mathletters}
%
%
\begin{table}
\caption{The $B\to K^{(*)}$ and $B\to \p(\rho)$ form factors of 
Melikhov and Nikitin \protect\cite{melikhov}.}\label{table:ff:mel}
\begin{tabular}{ccc}
Form factor & $B\to K$ & $B^-\to \p^-$  \\ \hline
$f_+(q^2)$ & $0.36 \left(1-\dis\frac{q^2}{6.88^2}\right)^{-2.32}$
& $0.29\left(1-\dis\frac{q^2}{6.71^2}\right)^{-2.35}$\\
$f_-(q^2)$ & $-0.30\left(1-\dis\frac{q^2}{6.71^2}\right)^{-2.27}$
& $-0.26\left(1-\dis\frac{q^2}{6.55^2}\right)^{-2.30}$\\
$s(q^2)$ & $0.06\ \GeV^{-1}\left(1-\dis\frac{q^2}{6.85^2}\right)^{-2.28}$
& $0.05\ \GeV^{-1}\left(1-\dis\frac{q^2}{6.68^2}\right)^{-2.31}$\\ 
\hline
Form factor & $B\to K^*$ & $B^-\to \rho^-$  \\ \hline
$g(q^2)$ & $0.048\ \GeV^{-1}\left(1-\dis\frac{q^2}{6.67^2}\right)^{-2.61}$
& $0.036\ \GeV^{-1}\left(1-\dis\frac{q^2}{6.55^2}\right)^{-2.75}$ \\
$f(q^2)$ & $1.61\ \GeV\left(1-\dis\frac{q^2}{5.86^2}+\dis\frac{q^4}{7.66^4}\right)^{-1}$ 
& $1.10\ \GeV\left(1-\dis\frac{q^2}{5.59^2}+\dis\frac{q^4}{7.10^4}\right)^{-1}$ \\ 
$a_+(q^2)$ & $-0.036\ \GeV^{-1}\left(1-\dis\frac{q^2}{7.33^2}\right)^{-2.85}$ 
& $-0.026\ \GeV^{-1}\left(1-\dis\frac{q^2}{7.29^2}\right)^{-3.04}$ \\
$a_-(q^2)$ & $0.041\ \GeV^{-1}\left(1-\dis\frac{q^2}{6.98^2}\right)^{-2.72}$ 
& $0.03\ \GeV^{-1}\left(1-\dis\frac{q^2}{6.88^2}\right)^{-2.85}$ \\

$h(q^2)$ & $0.0037\ \GeV^{-2}\left(1-\dis\frac{q^2}{6.57^2}\right)^{-3.28}$
& $0.003\ \GeV^{-2}\left(1-\dis\frac{q^2}{6.43^2}\right)^{-3.42}$ \\
$g_+(q^2)$ & $-0.28\left(1-\dis\frac{q^2}{6.67^2}\right)^{-2.62}$
& $-0.20\left(1-\dis\frac{q^2}{6.57^2}\right)^{-2.76}$ \\
$g_-(q^2)$ & $0.24\left(1-\dis\frac{q^2}{6.59^2}\right)^{-2.58}$
& $0.18\left(1-\dis\frac{q^2}{6.50^2}\right)^{-2.73}$
\end{tabular}
\end{table}
%
%
\begin{table}
\caption{The $B\to K^{(*)}$ form factors of Colangelo \ea\ \protect\cite{colangelo}, 
with $M=5\ \GeV$. As for the $B\to \p (\r)$ transition,  
we use the form factors listed below with $M=5.3\ \GeV$ in 
$F_1$ and $F_T$.}\label{table:ff:col}
\begin{tabular}{cc}
Form factor & $B\to K$ \\ \hline
$F_1(q^2)$  &$0.25\left(1-\dis\frac{q^2}{M^2}\right)^{-1}$\\
$F_0(q^2)$  &$0.25\left(1-\dis\frac{q^2}{49}\right)^{-1}$\\
$F_T(q^2)$  &$-0.14\left(1-\dis\frac{q^2}{M^2}\right)^{-1}
\left(1-\dis\frac{q^2}{49}\right)^{-1}$\\ 
\hline
Form factor &$B\to K^*$ \\ \hline
$V(q^2)$    &$0.47\left(1-\dis\frac{q^2}{25}\right)^{-1}$\\
$A_1(q^2)$  &$0.37(1-0.023 q^2)$\\
$A_2(q^2)$  &$0.40(1+0.034q^2)$\\
$A_0(q^2)$  &$0.30\left(1-\dis\frac{q^2}{4.8^2}\right)^{-1}$\\
$T_1(q^2)$  &$0.19\left(1-\dis\frac{q^2}{5.3^2}\right)^{-1}$\\
$T_2(q^2)$  &$0.19(1-0.02 q^2)$\\
$T_3(q^2)$  &$0.30(1+0.01 q^2)$
\end{tabular}
\end{table}
%
%


\begin{references}
\bibitem{ckm}N.~Cabibbo, \prl{10}{63}{531}; M.~Kobayashi and K.~Maskawa,
\ptp{49}{73}{652}.
\bibitem{epsilonprime}For recent reviews, see S.~Bertolini, J. O.~Eeg, and
M.~Fabbrichesi, hep-ph/0002234, and references therein; 
A. J. Buras, hep-ph/9908395, talk given at 
\emph{1999 Chicago Conference on Kaon Physics 
(KAON '99)}, Chicago, IL, 1999.
\bibitem{baryogenisis}A. G. Cohen, D. B. Kaplan, and 
A. E. Nelson, \ann{43}{93}{27}; A. D. Dolgov, hep-ph/9707419;
V. A. Rubakov and M. E. Shaposhnikov, Usp. Fiz. Nauk {\bf 166}, 493 (1996)
[Phys. Usp. {\bf 39}, 461 (1996)]. 
\bibitem{ali:rev}A.~Ali, \acta{27}{96}{3529}. 
\bibitem{joanne}J. L.~Hewett and J. D.~Wells, \prd{55}{97}{5549}.
\bibitem{susy:fcnc}M.~Misiak, S.~Pokorski, and J.~Rosiek,  
in \emph{Heavy Flavours II}, edited by A. J.~Buras and M.~Lindner 
(World Scientific, Singapore, 1998), p.~795, hep-ph/9703442.
\bibitem{mssm:original}H. P.~Nilles, \prp{110}{84}{1};
H. E.~Haber and G. L.~Kane, \ibid{117}{85}{75};
J. F.~Gunion and H. E.~Haber, \np{272}{86}{1}; \ib{B402}{93}{567} (E);
J. F.~Gunion, H. E.~Haber, G. L.~Kane, and S.~Dawson,  
\emph{The Higgs Hunter's Guide} (Addison-Wesley, Reading, MA, 1990).  
\bibitem{mssmrev}For recent reviews, see M.~Kuroda, hep-ph/9902340; 
J. C.~Rom\~{a}o, hep-ph/9811454.
\bibitem{bertolini:etal}S.~Bertolini, F.~Borzumati, A.~Masiero, and 
G.~Ridolfi, \np{353}{91}{591}.
\bibitem{ali:etal}A.~Ali, G. F.~Giudice, and T.~Mannel, \zpc{67}{95}{417}.
\bibitem{cho:etal}P.~Cho, M.~Misiak, and D.~Wyler, \prd{54}{96}{3329}.
\bibitem{goto:etal}T.~Goto, Y.~Okada, Y.~Shimizu, and M.~Tanaka, 
\prd{55}{97}{4273}; T.~Goto, Y.~Okada, and Y.~Shimizu, \ibid{58}{98}{094006}.
\bibitem{goto:etal:phase:sugra}T.~Goto, Y.-Y.~Keum, T.~Nihei, Y.~Okada, 
and Y.~Shimizu, \pl{460}{99}{333}.
\bibitem{off-diagonal}Y. G.~Kim, P.~Ko, and J. S.~Lee, \np{544}{99}{64}.
\bibitem{baek:ko}S.~Baek and P.~Ko, \prl{83}{99}{488}; \pl{462}{99}{95}, 
and references therein.
\bibitem{huang:etal}C.-S.~Huang, W.~Liao, and Q.-S.~Yan, \prd{59}{99}{011701};
C.-S.~Huang and Q.-S.~Yan, \pl{442}{98}{209}.  
\bibitem{lunghi:etal}E.~Lunghi, A.~Masiero, I.~Scimemi, and L.~Silvestrini, 
\npn{568}{00}{120}.
\bibitem{np1:exc}See, for example, C.~Greub, A.~Ioannissian, and D.~Wyler, 
\pl{346}{95}{149}; G.~Burdman, \prd{52}{95}{6400}; \ib{57}{98}{4254};
D.~Melikhov, N.~Nikitin, and S.~Simula, \pl{442}{98}{381};
T. M.~Aliev and M.~Savc{\i}, \prd{60}{99}{014005}.
\bibitem{np2:exc}A.~Ali, P.~Ball, L. T.~Handoko, and G.~Hiller, 
\prdn{61}{00}{074024}, and references therein. 
\bibitem{enrico:recent}E.~Lunghi and I.~Scimemi, \npn{574}{00}{43}.
\bibitem{four-body}F.~Kr\"uger, L. M.~Sehgal, N.~Sinha, and R.~Sinha, 
\prdn{61}{00}{114028}. 
\bibitem{fklms:exc}F.~Kr\"uger and L. M.~Sehgal, \prd{56}{97}{5452}; 
\ib{60}{99}{099905} (E).
\bibitem{phase:canc}T.~Ibrahim and P.~Nath, \prd{57}{98}{478}; 
\ib{58}{98}{019901} (E); \ib{58}{98}{111301}; 
M.~Brhlik, G. J.~Good, and G. L.~Kane, \ibid{59}{99}{115004};
A.~Bartl, T.~Gajdosik, W.~Porod, P.~Stockinger, and H.~Stremnitzer, 
\ibid{60}{99}{073003}; S.~Pokorski, J.~Rosiek, and C. A.~Savoy, 
\npn{570}{00}{81}.
\bibitem{phase:non-uni}S. A.~Abel and J. M.~Fr\`ere, \prd{55}{97}{1623}.
\bibitem{effsusy}A. G.~Cohen, D. B.~Kaplan, and A. E.~Nelson, 
\pl{388}{96}{588}; A.~Pomarol and D.~Tommasini, \np{466}{96}{3};
A. G.~Cohen, D. B.~Kaplan, F.~Lepeintre, and A. E.~Nelson, 
\prl{78}{97}{2300}.
\bibitem{mssmrev:romao}See the second paper in \rf{mssmrev}.
\bibitem{veff:one-loop}A.~Pilaftsis, \pl{435}{98}{88}; \prd{58}{98}{096010};
M.~Brhlik and G. L.~Kane, \pl{437}{98}{331}; D. A.~Demir, 
\prd{60}{99}{055006}; \ib{60}{99}{095007}; 
A.~Pilaftsis and C. E. M.~Wagner, \np{553}{99}{3}. 
\bibitem{phase:sugra}T.~Falk and K. A.~Olive, \pl{439}{98}{71};
E.~Accomando, R.~Arnowitt, and B.~Dutta, Phys. Rev. D 
(to be published), hep-ph/9907446.
\bibitem{offdiag:cmssm}D. A.~Demir, A.~Masiero, and O.~Vives, 
\prdn{61}{00}{075009}, and references therein.
\bibitem{edm:two-loop}D.~Chang, W.-Y.~Keung, and A.~Pilaftsis, 
\prl{82}{99}{900}; \ib{83}{99}{3972} (E).
\bibitem{bmm}A. J.~Buras and M.~M\"unz, \prd{52}{95}{186}; M.~Misiak, \np{393}{93}{23}; \ib{B439}{95}{461} (E).
\bibitem{review}For a review, see G.~Buchalla, A. J.~Buras, and 
M. E.~Lautenbacher, \rmp{68}{96}{1125}.
\bibitem{ckm:wolfenstein}L.~Wolfenstein, \prl{51}{83}{1945}.
\bibitem{ckm:analysis}F.~Parodi, P.~Rodeau, and A.~Stocchi, 
\nc{112}{99}{833}; A.~Ali and D.~London, \euro{9}{99}{687}.
\bibitem{ops:new}D.~Guetta and E.~Nardi, \prd{58}{98}{012001}; 
T. G.~Rizzo, \ibid{58}{98}{114014}; S.~Fukae, C. S. Kim, T.~Morozumi, and 
T.~Yoshikawa, \ibid{59}{99}{074013}. 
\bibitem{ali:hiller}A.~Ali and G.~Hiller, \prd{60}{99}{034017}.
\bibitem{fklms:res}F.~Kr\"uger and L. M.~Sehgal, \pl{380}{96}{199};
\prd{55}{97}{2799}.
\bibitem{dispersive}N.~Cabibbo and R.~Gatto, \pr{124}{61}{1577}.
\bibitem{burkhardt}H.~Burkhardt and B.~Pietrzyk, \pl{356}{95}{398}.
\bibitem{amm}A.~Ali, T.~Mannel, and T.~Morozumi, \pl{273}{91}{505}.
\bibitem{lw}Z.~Ligeti and M. B.~Wise, \prd{53}{96}{4937}; see also 
Z.~Ligeti, I. W.~Stewart, and M. B.~Wise, \pl{420}{98}{359}.
\bibitem{non-fac:review}F. M.~Al-Shamali and A. N.~Kamal, 
\prd{60}{99}{114019}, and references therein.
%
%
\bibitem{melikhov}D.~Melikhov and N.~Nikitin, hep-ph/9609503.
\bibitem{pdg}Particle Data Group, C.~Caso \ea, \euro{3}{98}{1}.
\bibitem{gounsak} T.~Kinoshita, B.~Ni\v{z}i\'c, and Y.~Okamata, \prd{31}{85}{2108}; F.~Jegerlehner, \zpc{32}{86}{195}.
\bibitem{colangelo}P.~Colangelo, F.~De Fazio, P.~Santorelli, and E.~Scrimieri, \prd{53}{96}{3672}; \ib{57}{98}{3186} (E).
\bibitem{melikhov:formulae}D.~Melikhov, N.~Nikitin, and  S.~Simula, 
\prd{57}{98}{6814}.
\bibitem{exp:btosg}ALEPH Collaboration, R.~Barate \ea, \pl{429}{98}{169}; 
CLEO Collaboration, S.~Ahmed \ea, hep-ex/9908022; see also G.~Eigen, 
hep-ex/9901005, talk given at \emph{4th International Symposium on 
Radiative Corrections (RADCOR '98)}, Barcelona, 1998.
\bibitem{kagan:neubert:btosg}A. L.~Kagan and M.~Neubert, \euro{7}{99}{5}. 
\bibitem{limits:exc:exp}CDF Collaboration, T.~Affolder \ea, 
\prl{83}{99}{3378}.
\bibitem{SMprediction}A.~Ali, \npps{59}{97}{86}.
\bibitem{mel:np}D.~Melikhov, N.~Nikitin, and S.~Simula, \pl{442}{98}{381}.
\bibitem{flavour-structure}M. Brhlik, L. Everett, 
G. L. Kane, S. F. King, and O. Lebedev, \prln{84}{00}{3041}; 
D. A. Demir, A. Masiero, and O. Vives, hep-ph/9911337.
\bibitem{kagan:neubert:cp}A. L.~Kagan and M.~Neubert, \prd{58}{98}{094012}.
\bibitem{ciuchini:etal}M.~Ciuchini, G.~Degrassi, P.~Gambino, and 
G. F.~Giudice, \np{527}{98}{21}; \ib{B534}{98}{3}; C.~Bobeth, M.~Misiak, 
and J.~Urban, \ibidn{B567}{00}{153}.
\end{references}
\end{document}